\documentclass[nofootinbib,prd,preprintnumbers,superscriptaddress,showpacs]{revtex4}

\pdfoutput=1
\usepackage{graphicx}
\usepackage{bm}
\usepackage{amssymb}
\usepackage{float}
\usepackage{amsmath}
\usepackage{subfigure}
\usepackage{dcolumn}
\usepackage{cancel}
\usepackage[colorlinks]{hyperref}
\usepackage[usenames,dvipsnames]{color}
\hypersetup{
     breaklinks=true,
    pdfstartview={FitH},    % fits the width of the page to the window
    colorlinks=true,       % false: boxed links; true: colored links
    linkcolor=blue,          % color of internal links
    citecolor=red,        % color of links to bibliography
    filecolor=magenta,      % color of file links
    urlcolor=blue,           % color of external links
    anchorcolor=green,      % Color for anchor text
    linktocpage=true
}

\def\doi{http://doi.org}

\begin{document}
\newcommand\be{\begin{equation}}
\newcommand\ee{\end{equation}}
\newcommand\bea{\begin{eqnarray}}
\newcommand\eea{\end{eqnarray}}
\newcommand\bseq{\begin{subequations}} %solo con amsmath
\newcommand\eseq{\end{subequations}}
\newcommand\bcas{\begin{cases}}
\newcommand\ecas{\end{cases}}
\newcommand{\p}{\partial}
\newcommand{\f}{\frac}

\title{Thermal phase transition in $F(R)$-charged $AdS_{4}$-scalar theory   }
%Einstein-conformally Maxwell invariant gravity}

\author {A. Rahmani}\email{a$_$rahmani@sbu.ac.ir}
\affiliation{Department of Physics, Shahid Beheshti University, G.C., Evin, Tehran 19839, Iran}

\author {M. Honardoost}\email{m$_$honardoost@sbu.ac.ir}
\affiliation{Department of Physics, Shahid Beheshti University, G.C., Evin, Tehran 19839, Iran}

\author {H. R. Sepangi}\email{hr-sepangi@sbu.ac.ir }
\affiliation{Department of Physics, Shahid Beheshti University, G.C., Evin, Tehran 19839, Iran}

%\date{\today}
 %---------------------------------------

\begin{abstract}

We investigate instabilities of $F(R)$-charged $AdS_{4}$ black holes by a massive charged scalar field in a linear perturbation regime. We study tachyonic instabilities as the near horizon scalar condensation in a model of $F(R)$  gravity with planar horizon and investigate properties of possible phase transitions. The results show that such transitions are sensitive to the first derivative of $F(R)$ with respect to $R$ in that the larger its value, the higher the critical temperature, thus resulting in a new generation of high-temperature superconductors. Also, for a certain range of parameters, $F(R)$-charged $AdS_{4}$ black holes suffer from superradiant instability. We consider the effects of the scalar mass and charge on such instabilities and conclude that RN black holes decay into  small hairy black holes that have a charged scalar condensate floating near the horizon. It is shown that the existence of phase transition at the critical temperature leading to a hairy black hole solution emerges for $T<T_{c}$, while RN black holes exist for $T>T_{c}$. The effect of $F(R)$ on the critical temperature is subsequently investigated in the case of superradiant instability, showing that higher critical temperatures would be possible in $F(R)$  gravity. We also check the stability of hairy black holes and show that the resulting hairy solution can be considered as a possible end point of superradiant instability of a small charged black hole.
\end{abstract}

\maketitle
\section{Introduction}
Black holes are one of the most interesting predictions of Einstein's General Relativity (GR) \cite{jac,herd1}. The historical LIGO direct detection of gravitational waves in 2016 \cite{ab} and the recent event horizon scale images of a supermassive black hole \cite{a1}, present strong evidence for the existence of astrophysical black holes.

A question that would come to mind regarding black holes is how stable such objects are? Technically, investigation of a black hole stability is an arduous problem due to non-linear partial differential equations resulting from field equations. However, recent improvements in numerical methods  have revived the issue once more and attracted considerable attentions in the context of alternative theories of gravity \cite{win, bosch, ra}.

In the past few decades, modified theories of gravity, generically known as $F(R)$ gravity with roots in string theory have emerged and used \cite{sol, sol1, hendi,hair} to explain  cosmological observations such as late time cosmic acceleration of the universe \cite{noj} and the existence of  dark matter/energy \cite{lis, sah} together with some theoretical challenging problems related to quantum gravity \cite{thi}. Among possible alternatives, the family of $F(R)$ gravity theories where $F$ is an arbitrary function of the Ricci scalar $R$,  nicely provides the freedom to explain late time cosmic accelerated expansion and structure formation of the universe without the need for mysterious dark matter/ energy \cite{noji, noji1, tsu}.
In addition, it was shown that imposing two conditions, namely $\frac{dF}{dR}>0$ and $\frac{d^{2}F}{dR^{2}}>0$ guarantee the existence of a well behaved $F(R)$ theory \cite{del}. Black hole solutions in $F(R)$ context have been widely studied in the past \cite{bhinfr, soti}. Interestingly, in the case of constant Ricci scalar, GR solutions are recovered and coincide with  vacuum solutions for $R_{0}=0$, $R_{0}>0$ and $R_{0}<0$, corresponding to asymptotically flat, de Sitter($dS$) and  Anti-de Sitter($AdS$)  solutions respectively \cite{hendi, hair}.

An interesting question regarding black hole solutions in such theories is the stability against various perturbations, specially in the case where there is no plain uniqueness theorem, e.g. in higher dimensions, which makes selecting physical solutions amongst a variety of possible solutions all the more interesting. A mechanism that would influence the stability of such objects is the process of energy extraction from a black hole. In particular, through superradiance phenomena, energy is extracted from rotating or charged black holes by scattering of a test scalar field  off it \cite{pani, brito}. In a confined system the scattered test particle will travel back and forth between regions near the event horizon and  boundary of the system, creating considerable back-reaction to the background and leads to the so-called \textit{Superradiant Instability}, see \cite{pani, press} for more details and discussions.
Although the scattered scalar field can be superradiantly amplified under specific conditions, it is clear that the \textit{instability} would occur only in an enclosed system. An artificial mirror-like boundary \cite{win, pani} as well as some intrinsic properties of the system such as a natural boundary in an $AdS$ space-time \cite{ads1, oscar1} or a massive scalar test particle \cite{hod1} can define a efficient confined system\footnote{It was shown in \cite{masswin, massherd} that contrary to the Kerr space-time, massive test scalar fields cannot trigger superradiant instability in the case of charged black holes.}. The question then arises as to what is expected as the final fate of an unstable black hole? A stable hairy black hole \cite{win, masswin, bosch} or an explosion event called bosenova \cite{yosh, herd} are two possible candidates describing the final state of a superradiantly unstable system both in GR or alternatively in $F(R)$ theories of gravity \cite{ra}.

One may also look at  black hole instabilities in the context of Anti-de Sitter/Conformal Field Theory (AdS/CFT) correspondence. The AdS/CFT correspondence is a formidable tool for analyzing strongly coupled gauge theories using classical gravitation \cite{holo, muss}. The duality is originated from string theory and  covers a wide range of applications in QCD, nuclear physics, non-equilibrium physics and condensed matter physics \cite{pen, 1-s2}. According to AdS/CFT duality, a static black hole in an $AdS$ space-time corresponds to a thermal state in CFT on the boundary \cite{hart, van}. Since perturbing a black hole in $AdS$ space-times corresponds to perturbing a thermal state in the CFT sector, the decay of perturbations is equivalent to  thermal equilibrium of such states \cite{hart1, gar}.
AdS/CFT duality is nicely used in condensed matter physics to envisage a dual gravity of strongly-coupled systems known as a high critical temperature superconductor. In this framework, since entropy must be continuous through second order phase transition, transition from a black hole to another state must occur in an $AdS$ space-time. It has been argued \cite{gub} that scalar-tensor theories having a charged scalar field around a charged black hole in $AdS$, would show transition from the initial black hole to a new hairy black hole. This means that in scalar-tensor theories, the initial black hole state may suffer from an instability so that a scalar hairy black hole emerges as the final state.

In this paper we consider a charged black hole with a scalar field around it and study the effects of two possible instabilities; near horizon scalar condensation \cite{hart1,  gar, gub, oscar} and superradiant instability. The latter is presented only in a global $AdS$ space-time and requires both the scalar field and black hole be charged. Alternatively, near horizon condensation instability (and the associated hairy black hole) was first found in a local planar $AdS$ \cite{oscar} and opened a plethora of research in the field of \textit{`` holographic super conductivity."}

The organization of the paper is as follows. In section \ref{BHa} we study near horizon scalar condensation instability in $F(R)$ gravity and discuss  the effects of $F(R)$ on critical temperature $T_{c}$. Our results show that $T_{c}$ will increase with the growth of $F_R=\frac{dF}{dR}$ so that a new generation of high-temperature super conductors can be introduced in the CFT counterpart. In section \ref{BHa1} superradiant instability of small $AdS$ black holes is considered and the effects of $F(R)$ on thermal transition studied. Conclusions are drawn in section \ref{DC}.

\section{New generation of high-temperature superconductors emerging from $F(R)$ gravity  \label{BHa}}
In 1950, Landau and Ginzburg presented a new description of superconductivity based on a second order phase transition where the order parameter is a complex scalar field $\phi$ \cite{Gin}.
Ignoring higher powers of $\phi$, the free energy $\cal F$ is assumed to take the form
\begin{equation}\label{1}
  {\cal F} = a(T-T_{c})|\phi|^{2}+\frac{b}{2}|\phi|^{4},
\end{equation}
where $a$ and $b$ are two positive constants. For $T>T_{c}$, the minimum of $\cal F$ occurs at $\phi=0$ while ${\cal F}_{min}$ takes a non-zero value for $\phi$ when $T<T_{c}$, reminiscent of the  Higgs mechanism in particle physics. In 1957, BCS theory was presented by Bardeen, Cooper and Schrieffer, bringing superconductivity on a more complete footing \cite{bar}. The theory is based on forming Cooper-pairs which condensate through a second order phase transition below a critical temperature $T_{c}$. The discovery of a new class of high $T_{c}$ superconductors in 1986 \cite{bed} with $T_{c}=134 K$ at atmospheric pressure demands a new description of superconductivity. Although there is evidence that electron pairs can still form in such high temperatures, the pairing mechanism is unclear. Fortunately these strongly coupled field theories can be studied in the context of AdS/CFT duality. The \textit{``holographic superconductors''} represent theories who have dual counterparts in gravity. It has been argued \cite{gub}  that a charged scalar field around a charged black hole in an $AdS$ framework can be the gravity part of a holographic superconductor. The action is
\begin{equation}\label{2}
  S=\frac{1}{2}\int d^{4}x \sqrt{-g} \left[R+\frac{6}{L^{2}}-\frac{1}{4}F_{ab}F^{ab} -g^{ab}D_{(a}^* \psi^* D_{b)} \psi-m_{0}^2 \psi \psi^*\right],
\end{equation}
where $F_{ab}=\nabla _{a} A_{b}-\nabla_{b}A_{a}$ is the electromagnetic field, $A_{a}$ the vector potential, $D_{a}=\nabla_{a}-iqA_{a}$ and $\psi$ a complex scalar field. The asterisk *, $m_{0}$ and $q$ signify complex conjugate, the mass and charge of the scalar field and $8\pi G=C=1$. The asymptotically Anti-de Sitter Reissner-Nordstr\"{o}m (RN) black hole and trivial scalar field $\psi=$constant is a solution of action (\ref{2}). However, this solution is not thermodynamically favored at very low temperatures \cite{gar}, in other words, if one starts from asymptotically {$AdS$}-RN solution and gradually lowers the temperature, one faces instability in the background. The possible hairy solution as the final state of instability is what one needs in the framework of holographic superconductivity \cite{hart1, gar}. As was mentioned in the introduction, for the theory presented in (\ref{2}), two different types of linear instability would occur. The first  is near horizon condensation and the second is known as superradiant instability.  In what follows, we investigate these instabilities in $F(R)$ gravity and start with the former type and continue to study superradiant instability in section \ref{BHa1}. To do so, we take the following $F(R)$
	\begin{equation}\label{model}
	F(R)=R-\frac{\alpha c_1 \left(\frac{R}{\alpha}\right)^n}{1+\beta\left(\frac{R}{\alpha}\right)^n}.
	\end{equation}
and consider black hole solutions with constant negative curvature in order to have a topological Schwarzschild-$AdS$ space-time \cite{hendi} which creates an infinite potential wall at the $AdS$ boundary and makes the system closed \cite{li}. The motivation behind the above choice is that it satisfies both cosmological and solar system experiments \cite{hu, nature, moon} and by setting the free parameters $\alpha=0.001$, $n=1$ and $\beta=3$, the model is free of ghost and tachyonic instability\footnote{ Using the condition  $\frac{dF}{dR}>0$ and $\frac{d^2 F}{dR^2}>0$ which is necessary to have a well-behaved theory, one can easily computes $c_1$. }.
\subsection{Near horizon condensation of planar black holes in $F(R)$ gravity \label{BHb}}
We start with the action of scalar-tensor $F(R)$ theory
\begin{equation}\label{3}
  S=\frac{1}{2}\int d^{4}x \sqrt{-g} \left[F(R)-\frac{1}{4}F_{ab}F^{ab} -g^{ab} D_{(a}^* \psi^* D_{b)} \psi-m_{0}^2 \psi \psi^*\right].
\end{equation}
Of course, an asymptotically {$AdS$}-RN black hole and a trivial scalar field are possible solutions of (\ref{3}), but it is the behaviour of the effective scalar mass that determines the stability of the system. Breitenl\"{o}hner and Freedman found that as long as $m^{2}_{eff}\gg m^{2}_{BF}$ where $m^{2}_{eff}\equiv -\frac{d^{2}}{4L^{2}}$, the scalar field in $AdS_{d+1}$ is normalizeable and that stable, asymptotically $AdS_{d+1}$ solutions in the UV region would exist. It was shown by Gubser and Hartnoll that planar, asymptotically $AdS$ black holes can become unstable when ``near horizon condensation'' mechanism is considered, see \cite{hart1, gub} for more details. Condensation starts near the horizon when the effective mass, $m^{2}_{eff}\equiv -\frac{1}{4L^{2}}$, violates the BF bound  of $AdS_{2}$ and forms charged scalar hair, hence providing all we need to describe a holographic superconductor in the CFT side. We follow this scenario for planar black holes in $F(R)$  gravity and trace the effects of $F(R)$ to critical temperature, $T_{c}$, which corresponds to the onset of phase transition in the CFT part.

Varying action (\ref{3}) results in
\begin{equation}\label{4}
 F_{R}R_{a b}-\frac{1}{2}F(R)g_{a b}-\nabla_{a}\nabla_{b}F_{R}+g_{a b}\Box F_{R} = kT_{a b} ,
\end{equation}
\begin{eqnarray}
   && \nabla_{a}F^{ab}=J^{b}, \label{5}\\
   &&D_{a}D^{a}\psi -m_{0}^2 \psi=0,\label{6}\\
   &&T_{ab}=T_{ab}^{\psi}+T_{ab}^{F},\label{7}
\end{eqnarray}
where the current $J^{a}$ and energy-momentum tensor $T_{ab}$ are given by
\begin{eqnarray}
   && J^{a}=\frac{iq}{2}[\psi^{*}D^{a}\psi-\psi(D^{a}\psi)^{*}], \label{8}\\
   &&T_{ab}^{\psi}=D^{*}_{(a}\psi^{*}D_{b)}\psi-\frac{1}{2}g_{ab}[g^{cd}D^{*}_{(c}\psi^{*} D_{d)}\psi+m_{0}^{2}\psi \psi^{*}],\label{9}\\
   &&T_{ab}^{F}=F_{ac}F_{b}^{c}-\frac{1}{4}g_{ab}F_{cd}F^{cd}\label{10}.
\end{eqnarray}
We are interested in a static, spherically symmetric solution with a negative, constant Ricci scalar, that is $R_{0}<0$,  and start with the ansatz
\begin{equation}\label{11}
ds^{2} =-N(r)h(r)dt^{2}+ \frac{dr^{2}}{N(r)}+r^{2}d\Omega^{2},
\end{equation}
where  for black holes with a planar horizon, $N(r)$ is defined as
\begin{equation}\label{12}
  N(r)=-\frac{2M}{r}+\frac{Q^{2}}{F_{R}(R_{0})r^{2}}-\frac{R_{0}}{12}r^{2},
\end{equation}
and $h(r)$ represents back reaction of the scalar field with $d\Omega^2$ being the usual flat line element. In the case of a local $AdS$ solution, i.e. a planar black hole, $Q$  and $M$ are the charge and mass densities respectively \cite{gub}. It is worth noting that for planar solutions in the gravity part, the dual theory in the CFT side lives in a flat space \cite{bas}. Assuming $\psi = \psi(r)$ and $A(r)=\phi(r) dt$ as the gauge potential, four non-trivial equations are obtained  using (\ref{4}-\ref{7}) for $F(R)=R+f(R)$

\begin{equation}\label{13}
(1+f_{R}) h'(r)=\frac{r q^2 \phi(r)^2 \psi(r)^2}{N(r)^2}+r h(r) \psi'(r)^2,
\end{equation}
\begin{equation}\label{14}
\phi'(r)^2+m_{0}^2 h(r) \psi(r)^2=-\frac{2(1+f_{R})}{r}\left[h(r)\left(N'(r)+\frac{N(r)}{r}\right)+\frac{1}{2} N(r) h'(r)\right] -f_{R} h(r) R_{0}+f h(r),
\end{equation}
\begin{equation}\label{15}
N(r)\phi''(r)+\left(\frac{2N(r)}{r}-\frac{N(r) h'(r)}{2 h(r)}\right)\phi'(r)-q^2\psi(r)^2\phi(r)=0,
\end{equation}
\begin{equation}\label{16}
N(r)\psi''(r)+\left(\frac{2N(r)}{r}+N'(r)+\frac{N(r)h'(r)}{2h(r)}\right)\psi'(r)+\left(\frac{q^2 \phi(r)^2}{N(r)h(r)}-m_{0}^2\right)\psi(r)=0,
\end{equation}
where $f_{R}=\frac{df}{dR}$ and a prime signifies  derivative with respect to $r$. To verify the existence of hairy solutions we  integrate (\ref{13}-\ref{16}) numerically. For this purpose, we expand field variables in the neighborhood of the event horizon, $N=N'_{h}(r-r_{h})+...$, $h=h_{h}+h'_{h}(r-r_{h})+...$, $\phi=\phi_{h}+\phi'_{h}(r-r_{h})+...$,  $\psi=\psi_{h}+\psi'_{h}(r-r_{h})+...$ \cite{ra}, where the event horizon $r_{h}$ is the largest real root of $N(r)$, given by
\begin{equation}\label{horizon}
r_h=\frac{1}{2 \sqrt{3}}\sqrt{\frac{2\times 6^{2/3}F_{R_0} L^2Q^2+6^{1/3}a_1^{2/3}}{F_{R_0} a_1^{1/3}}}+\frac{1}{2}\sqrt{-\frac{2\times 2^{2/3} L^2 Q^2}{3^{1/3} a_1^{1/3}}-\frac{2^{1/3}a_1^{1/3}}{3^{2/3}F_{R_0}}+\frac{4 \sqrt{3}L^2 M}{\sqrt{\frac{2\times 6^{2/3}F_{R_0}L^2Q^2+6^{1/3}a_1^{2/3}}{F_{R_0} a_1^{1/3}}}}},
	\end{equation}	
and $L^2=-\frac{12}{R_0}$, $F_{R_0}\equiv F_R(R_0)$ with $a_1 \equiv 9 F_{R_0}^3L^4M^2+\sqrt{81 F_{R_0}^6L^8M^4-48 F_{R_0}^3L^6Q^6}$.

Equation (\ref{15}) indicates that in order to have regular field variables at the event horizon, one should have $\phi_{h}=0$. Plugging these expansions in (\ref{13}), (\ref{14}) and (\ref{16}), we get
\begin{eqnarray}
&&N'_{h}=\frac{r_{h}}{2(1+f_{R_{0}})}\left(\frac{{\phi'_{h}}^2}{h_{h}}+m_{0}^{2} \psi_{h}^2-f+f_{R_{0}}R_{0} \right),\label{17} \\
&&\psi'_{h}=\frac{2\left(1+f_{R_{0}}\right)m_{0}^2 \psi_{h}r_{h}}{-r_{h}^2\left(\frac{{\phi'_{h}}^2}{h_{h}}+m_{0}^{2} \psi_{h}^2-f+f_{R_{0}}R_{0} \right)+2\left(1+f_{R_{0}}\right)},\label{18}\\
&&h'_{h}=\frac{4r_{h}^{3}\psi_{h}^{2}\left(1+f_{R_{0}}\right)\left(q^2{\phi'_{h}}^{2}+m_{0}^{4}\right)}{\left[-r_{h}^2\left(\frac{{\phi'_{h}}^2}{h_{h}}+m_{0}^{2} \psi_{h}^{2}-f+f_{R_{0}}R_{0} \right)+2\left(1+f_{R_{0}}\right)\right]^2}.\label{19}
\end{eqnarray}
We now impose the reflective boundary condition as vanishing of the scalar field at the $AdS$ boundary. Since condensation occurs only near the event horizon, we need  $h(r\longrightarrow \infty)\equiv h_{\infty}=1$ which also guarantees that Hawking temperature of the black hole is equivalent to  temperature of the field theory at the boundary \cite{hart,bas}. From the general asymptotic behavior of the scalar field and vector potential \cite{hart}
\begin{eqnarray}
&&\psi=\frac{\psi_{1}}{r^{\Delta_{-}}}+\frac{\psi_{2}}{r^{\Delta_{+}}},\nonumber\\
&&\phi=\mu-\frac{\rho}{r}\label{19},
\end{eqnarray}
one may determine charge density of the black hole and expectation values of the scalar field where $\Delta_{\mp}=\frac{3}{2}\mp\sqrt{\frac{9}{4}+m_{0}^2L^2}$, $\psi_{1}$ and $\psi_{2}$ are the expectation values of the scalar field, $\mu$ is the chemical potential and $\rho$ is the charge density. It is well known that for having a stable theory we must
impose either $\psi_{1}$ or $\psi_{2}$ to vanish \cite{hart1}. Therefore, the expectation values of dual operator $\emph{O}_{\psi_{1}/\psi_{2}}$ is determined by the non-vanishing components. In this paper we assume the boundary conditions for the systems at $r\longrightarrow \infty$ to be $\psi_{1}=0$ which means that  $\emph{O}_{2}\varpropto \psi_{2} $. Note that we investigate the properties of dual field theory from the asymptotic behavior of solutions.

\subsection{Numerical solution and results\label{BH1a}}

To progress further, we use the shooting method for a fully numerical integration of the coupled equations (\ref{13}-\ref{16}) from $r_{h}+\epsilon$ to the reflective boundary. To avoid singular solutions at $r_{h}$, we take $r_{h}\longrightarrow r_{h}+\epsilon$ with $\epsilon=10^{-8}$. We also use the scaling symmetry of the equations of motion for further simplification \cite{oscar}
\begin{equation}\label{21}
(t,r,\theta,\varphi)\rightarrow (\lambda t,\lambda r,\theta,\varphi),\qquad(N,h,\phi,\psi)\rightarrow (N,h,\phi,\psi),\qquad (q,L,r_{h})\rightarrow (\frac{q}{\lambda},\lambda L,\lambda r_{h}),
\end{equation}
and thus fix $L=1$ and rescale all quantities to have dimensionless equations. Furthermore, there is an additional scaling symmetry for asymptotically local solutions, i.e. planar solutions, that allows us to fix $r_{h}=1$ too \cite{bas}.

\begin{figure}[!ht]
\includegraphics[width=8cm,height=5.5cm]{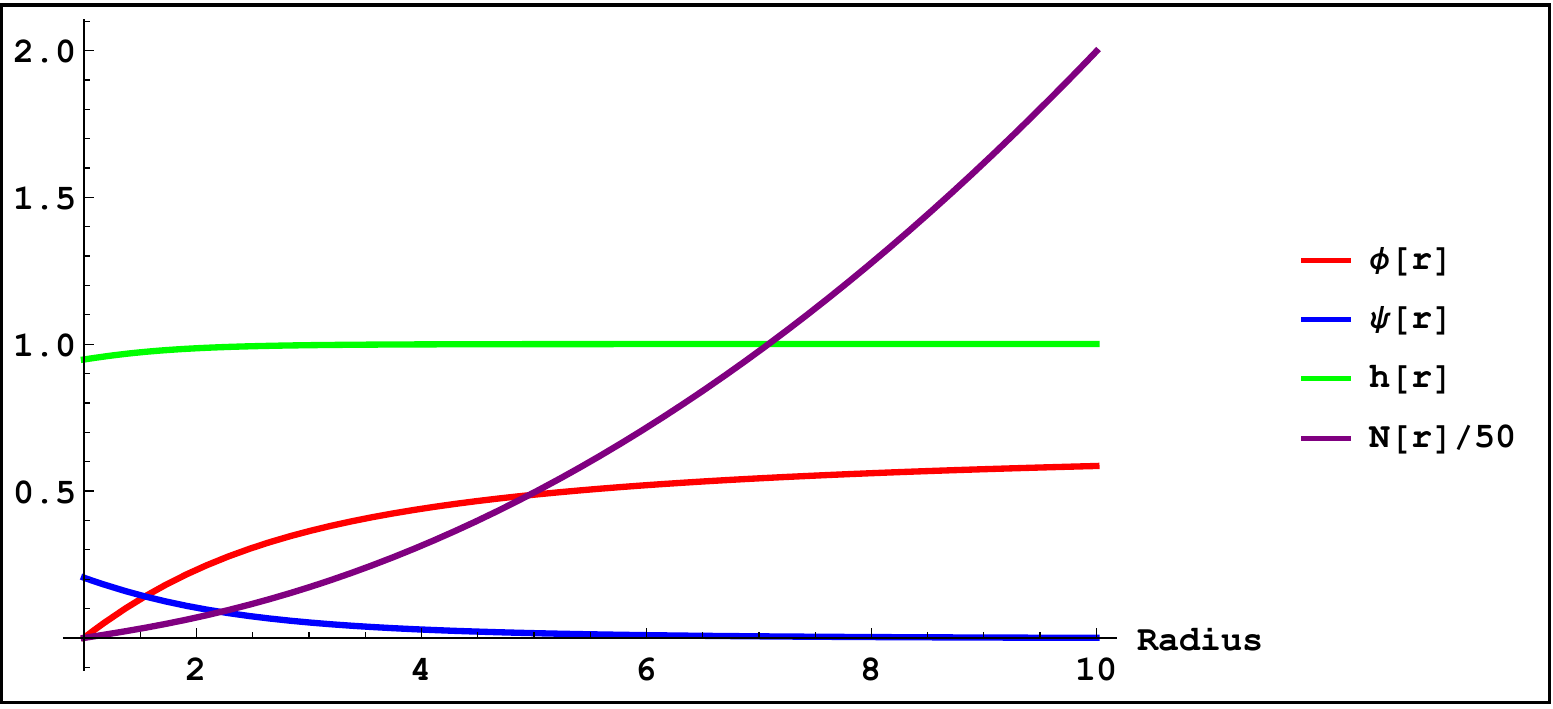}
\includegraphics[width=8cm,height=5.5cm]{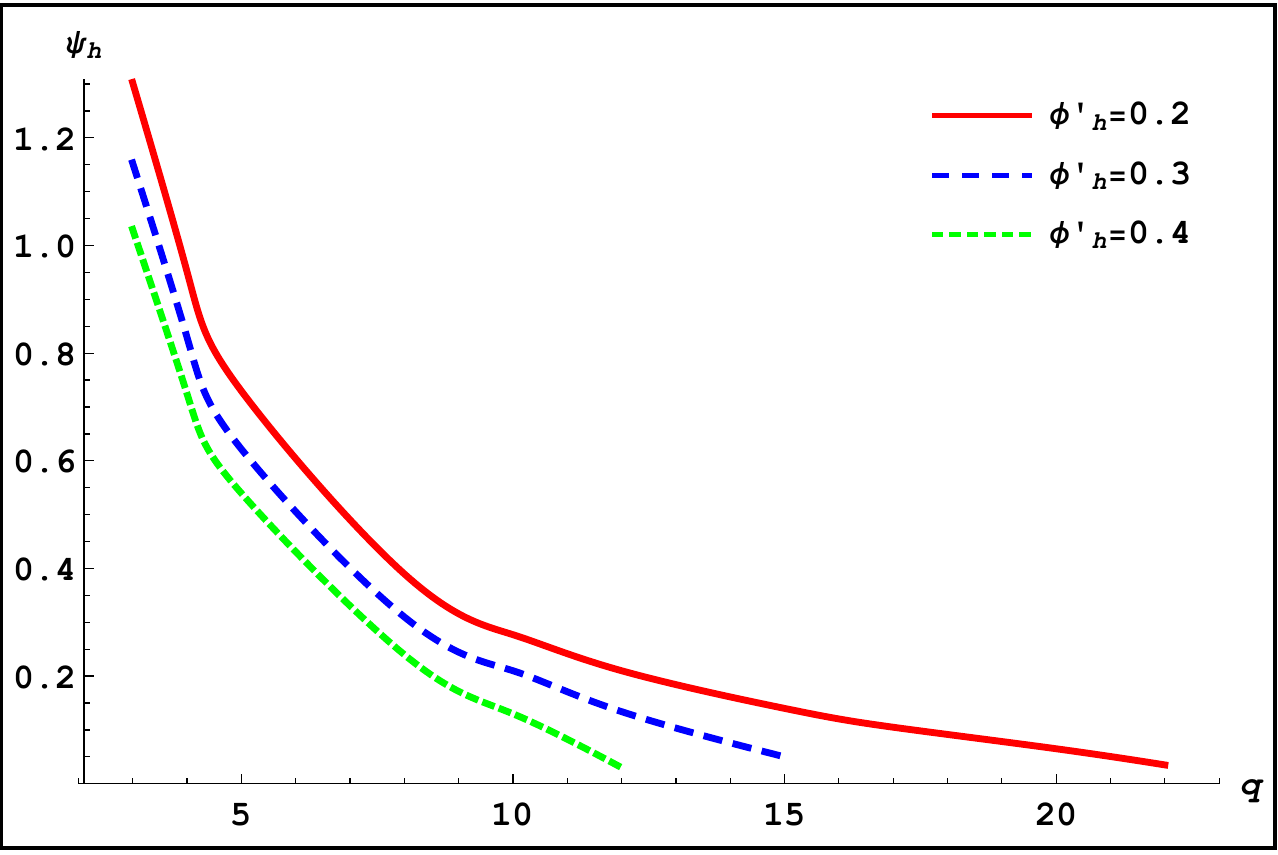}
\caption{Left: Plot of field variables as a function of radius with $q=10$, $m_{0}^{2}=-2$, $\psi_{h}=0.205$ and $\phi'_{h}=0.3$. Right: $\psi_{h}$ is plotted as a function of $q$ when the scalar field has only one node at the reflective boundary with $m_{0}^{2}=-2$ and different values of $\phi'_{h}$. }
\label{fig1p}
\end{figure}

\begin{figure}[!ht]
\includegraphics[width=8cm,height=5.5cm]{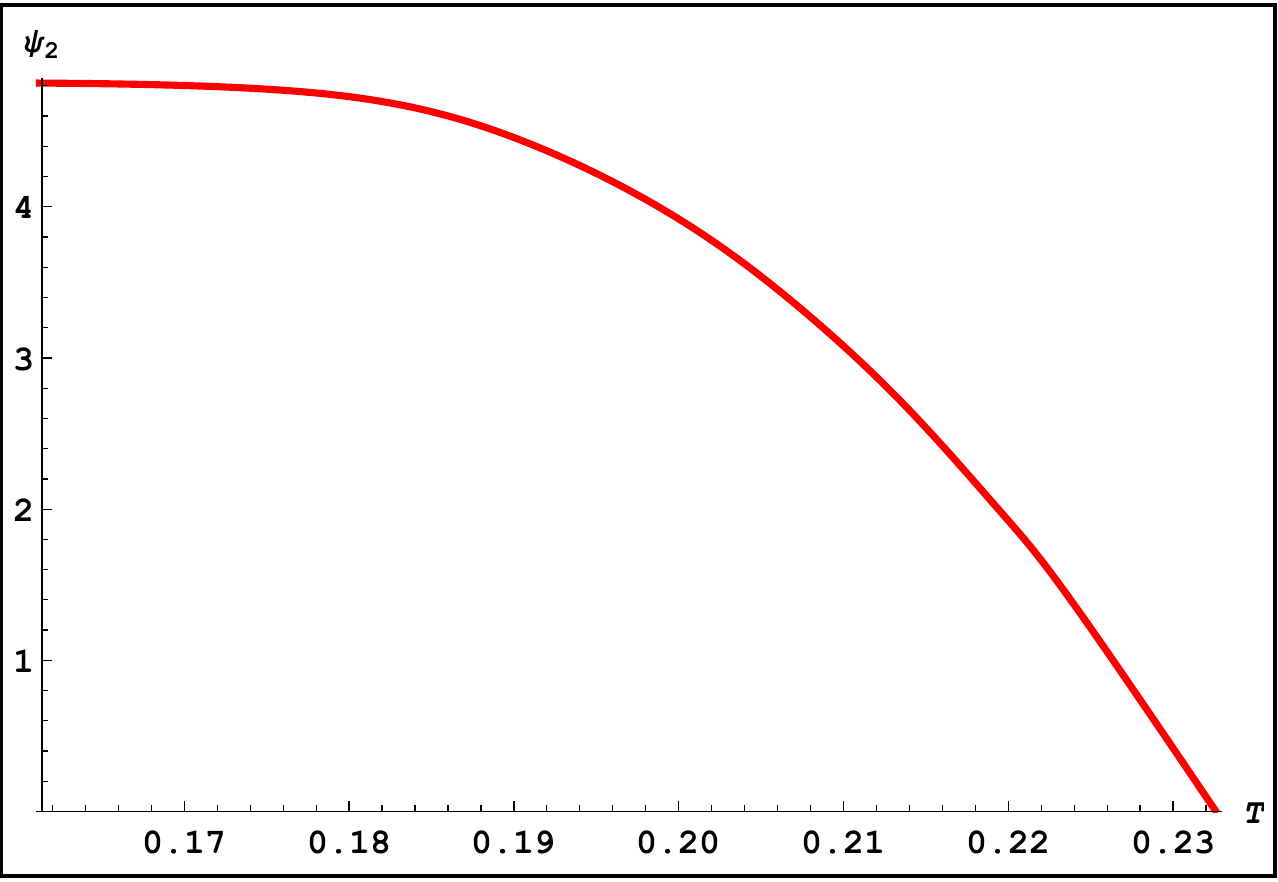}
\includegraphics[width=8cm,height=5.5cm]{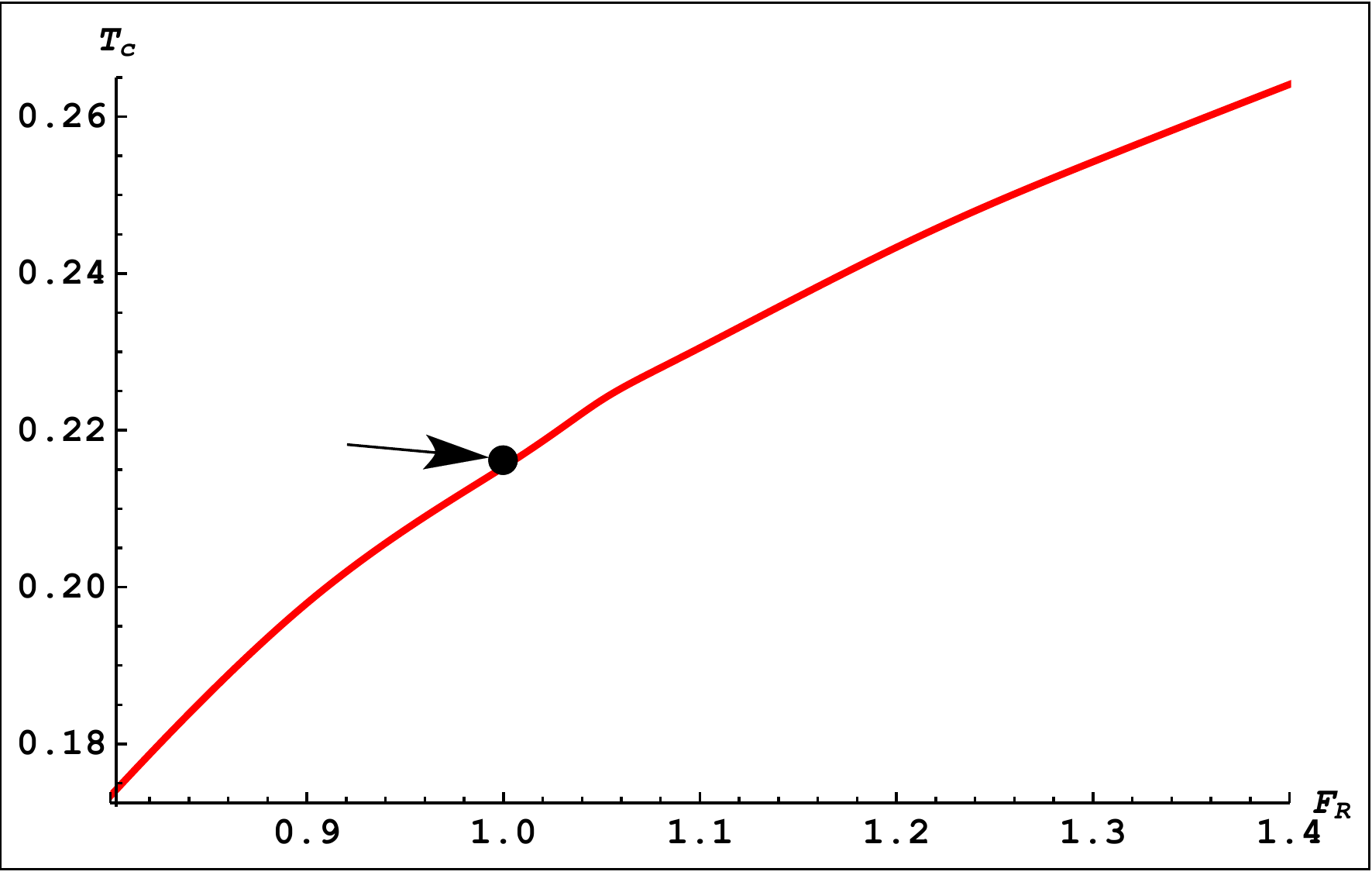}
\caption{Left: Behavior of the condensate as a function of temperature $T$ with $m_{0}^2=-2$ and $q=4$. Right: $F_{R}$ as a function of critical temperature for $q=5$ and $m_{0}^2=-2$. The arrow indicates critical temperature in $AdS$ Einstein-Maxwell scalar field theory. }
\label{fig2p}
\end{figure}

We now move on to investigate the existence of hairy black hole solutions and the possibility of phase transition from a charged $F(R)-AdS_{4}$ black hole to a hairy black hole for the model represented by (\ref{model}). The left panel in Fig. \ref{fig1p} confirms the existence of hairy solutions whereas the right panel shows the phase space ($\psi_{h}$-$q$) of solutions with only one node at the reflective boundary for different values of $\phi'_{h}$.
Using Hawking temperature for black holes $T_{H}=\frac{N'_{h}\sqrt{h_{h}}}{4 \pi}$, one can see the behaviour of the condensate in terms of critical temperature. The critical temperature $T_{c}$ is the point at which the scalar field appears as the zero mode, i.e. $\psi_{2}=0$, and the condensate $\psi_{2}$ is a parameter which explains properties of the phase transition. In the left panel of Fig. \ref{fig2p},  we consider behavior of the condensate as a function of $T$ by changing the initial conditions which shows that there is a condensate for $T<T_{c}$ and phase transition is of second order due to its compatibility with a square root law $\psi_{2}\varpropto (1-\frac{T}{T_{c}})^{\frac{1}{2}}$ \cite{wang}.
The right panel in Fig. \ref{fig2p} shows critical temperature in terms of $F_{R}$. It is clear that theories with higher values of $F_{R}$ describe a new generation of superconductors with higher critical temperature. The arrow in the right panel refers to the case for which $F_{R}=1$, corresponding to GR. This means that for certain choices of $F(R)$, as the one above, one attains a higher critical temperature $T_c$ relative to GR.

\section{Superradiant instability of small black holes in $F(R)$ gravity\label{BHa1}}
In the previous section we considered black holes with planar horizon which refer to local asymptotically $AdS$ space-times. In the following we consider
 \begin{equation}\label{22}
  N(r)=1-\frac{2M}{r}+\frac{Q^{2}}{F_{R}(R_{0})r^{2}}-\frac{R_{0}}{12}r^{2},
\end{equation}
which defines global asymptotically $AdS_{4}$ space-times provided $R_{0}=-\frac{12}{L^{2}}=4\Lambda$ \cite{ahmad, zha}. A global $AdS_{4}$ system would suffer from both near-horizon scalar condensation and superradiant instability \cite{oscar} simultaneously but for a certain limit one would dominate the other. It has been shown that large global {$AdS$}-RN black holes only have near horizon instability while superradiant instability exits just in small global {$AdS$}-RN black holes \cite{oscar}.

To study superradiant condition, we first neglect back reaction of the scalar field on background, supposing a monochromatic and spherically-symmetric perturbation with frequency $\omega$
\begin{equation}\label{23}
 \psi (r,t)= \frac{e^{-i\omega t}\Psi(r)}{r}.
\end{equation}
We take the vector potential as $A_{a}dx^{a}=\phi(r) dt =(-\frac{Q}{r}+c)dt$ where $c=\frac{Q}{r_{h}}$,  for which $AdS$ black holes have a vanishing vector potential on the event horizon in the context of  AdS/CFT correspondence \cite{uchi}. This integration constant only affects the value of the real part of frequency as $\mbox{Re}(\omega)\rightarrow \mbox{Re}(\omega)+ q c$ while has no effect on the imaginary part of frequency $\mbox{Im}(\omega)$. Using  (\ref{6}), the radial part of massive charged scalar perturbations is written as
\begin{equation}\label{24}
\frac{d^{2}\Psi}{dr_{*}^{2}}+V_{eff} \Psi=0,
\end{equation}
where $\frac{dr_{*}}{dr}=\frac{1}{N}$ and
\begin{eqnarray}\label{25}
V_{eff}=-N\left(\frac{l(l+1)}{r^{2}}+\frac{N'}{r}+m_{0}^2\right)+\left(\omega+q \phi\right)^2.
\end{eqnarray}
Using asymptotic behaviour of (\ref{24}), the radial solutions near horizon and at infinity are given by
\begin{eqnarray}\label{26}
&&\Psi \sim r^{-\frac{1}{2}\left(1+\sqrt{9+4m_{0}^2 L^2}\right)}         \qquad r_{*}\rightarrow +\infty ,\nonumber\\
 &&\Psi \sim e^{-i\omega r_{*}}                     \qquad r_{*}\rightarrow -\infty .
\end{eqnarray}
For a massive, charged scalar field in a small RN-{$AdS_4$} black hole, the real and imaginary parts of the quasinormal frequency for the lowest order modes have analytically been derived in \cite{her} and are given by
\begin{eqnarray}\label{27}
&&\mbox{Re}(\omega)=\frac{3}{2L}+\sqrt{m_{0}^2+\frac{9}{4 L^2}}-\frac{q Q}{r_{h}} ,\nonumber\\
 &&\mbox{Im}(\omega) = -2\frac{r_{h}^2 \times\Gamma\left(\frac{3}{2}+\sqrt{m_{0}^2L^2+\frac{9}{4}}\right)}{  \Gamma(\frac{1}{2})\times\Gamma\left(1+\sqrt{m_{0}^2L^2+\frac{9}{4}}\right)} \times \frac{\mbox{Re}(\omega)}{L^2} .
\end{eqnarray}
Since Klein-Gordon equation in $F(R)$ gravity with constant negative curvature is the same as that in an $AdS$ space-time, the results of \cite{uchi, her} are also valid for $F(R)$  and we use (\ref{27}) as the fundamental frequency. It is clear that $\mbox{Im}(\omega)>0$  which corresponds to $\mbox{Re}(\omega)<0$, shows the superradiance regime and leads to exponential growth of perturbations. As was mentioned before, the growth of perturbations in an effectively enclosed $AdS$ space-time results in an instability and can be best studied numerically using a shooting method, see \cite{ra} for more details.

 \begin{figure}[!ht]
\includegraphics[width=8cm,height=5.5cm]{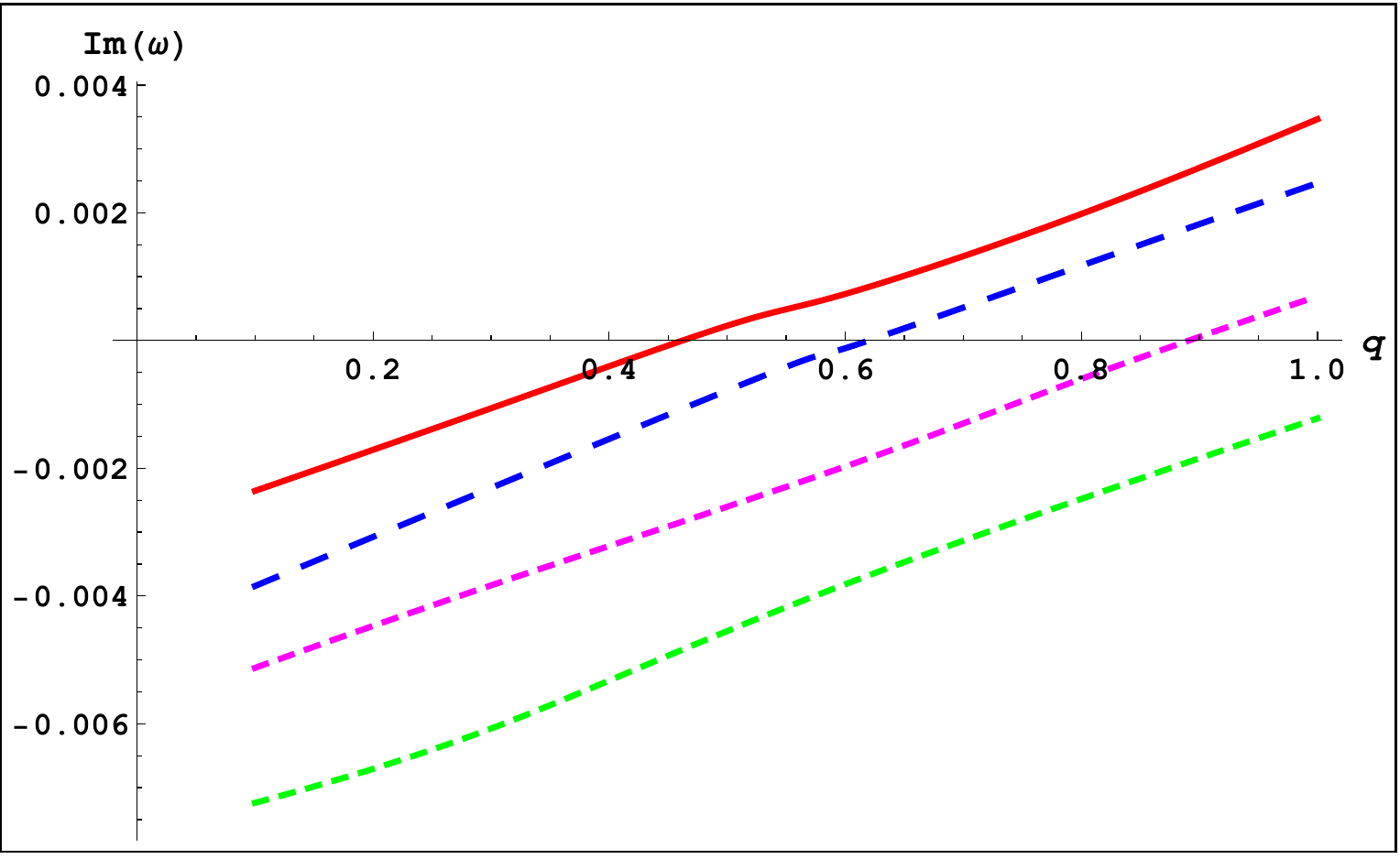}
\includegraphics[width=8cm,height=5.5cm]{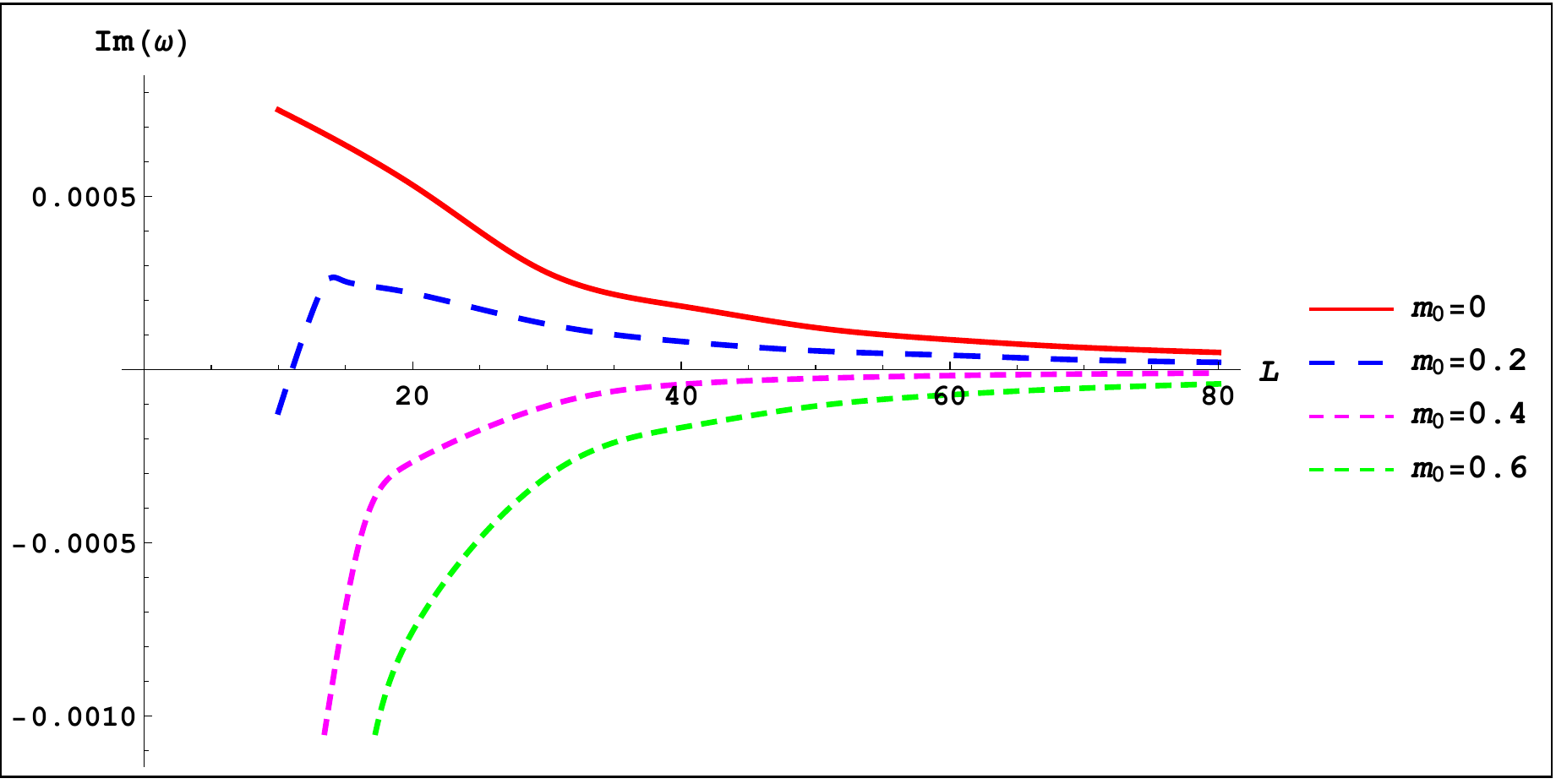}
\caption{Imaginary parts of the frequency for model (\ref{model}) with $\alpha=0.001$, $\beta=3$, $n=1$ and different values of the scalar mass as a function of, Left: the scalar charge for $L=10$, Right: $AdS$ radius for $q=0.6$  with $M=1$ and $Q=0.9$. The plots represent phase transition where the sign of the imaginary part of the frequency changes from negative to positive.  }
\label{fig3}
\end{figure}
Fig. \ref{fig3} shows the behaviour of $\mbox{Im}(\omega)$ for (\ref{model}) as a function of $q$ and $AdS$ radius for different values of the scalar field mass. As can be seen,  $\mbox{Im}(\omega)$ depends on $q$ and $m_{0}$. There is a direct relation between $\mbox{Im}(\omega)$ and $q$, i.e. superradiant instability becomes stronger by increasing $q$. The right panel shows that in the case of $q=m_{0}$ (we note that $m_0$ is considered as $m_0 \times L$ with $L=1$),   $\mbox{Im}(\omega)$ becomes negative and superradiant instability dose not occur while the instability may happen for $q>m_{0}$ and increases by decreasing $m_{0}$.

\begin{figure}[!ht]
	\includegraphics[width=10cm,height=5.5cm]{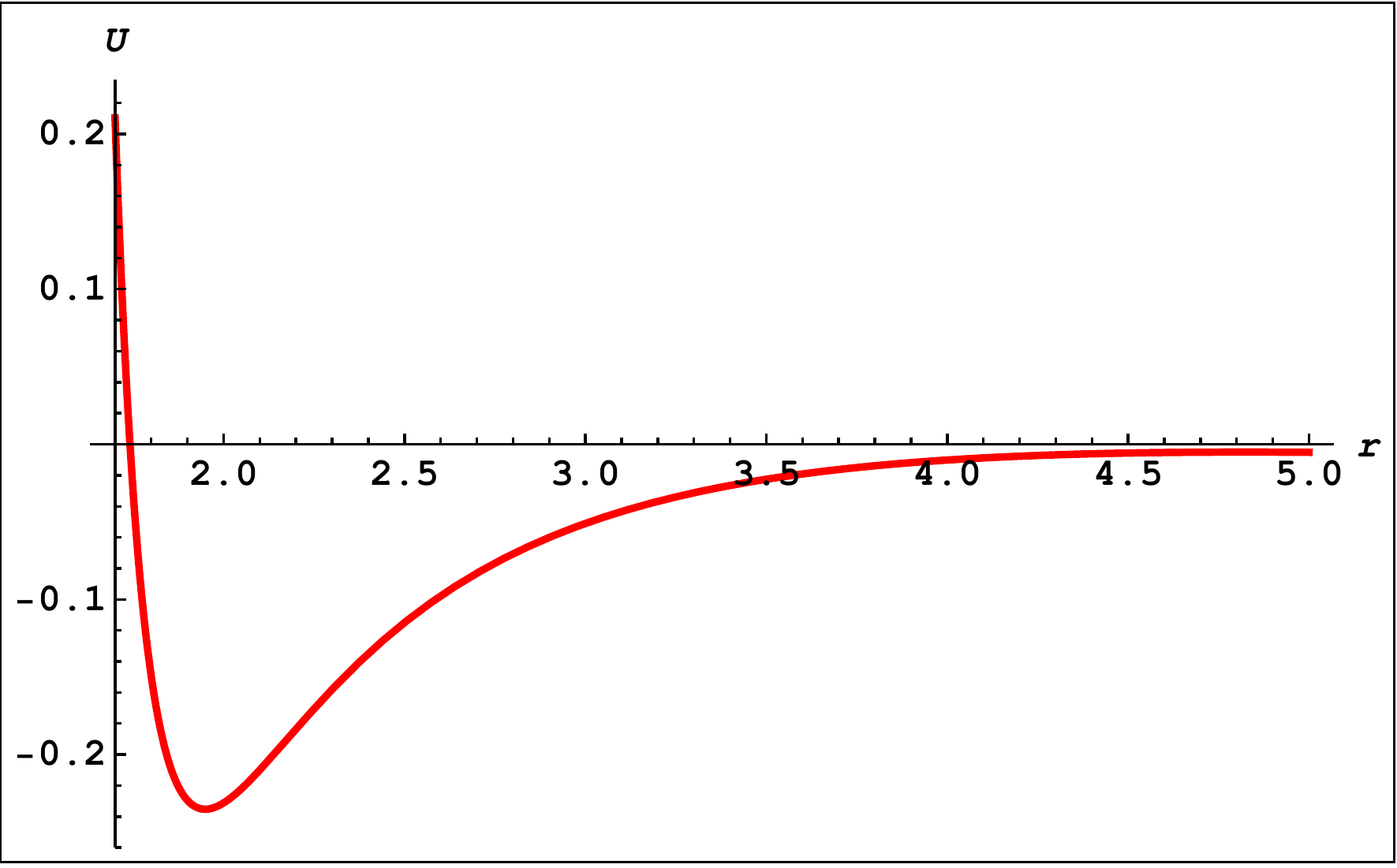}
	\caption{The shape of the radial Schrodinger potential $U$ as a function of $r$ for model (\ref{model}) with $\alpha=0.001$, $\beta=3$, $n=1$, $L=10$, $Q=0.9$, $M=1$, $q=1$ and $m_0=0.4$.  }
	\label{well}
\end{figure}

\begin{figure}[!ht]
	\includegraphics[width=8cm,height=5.5cm]{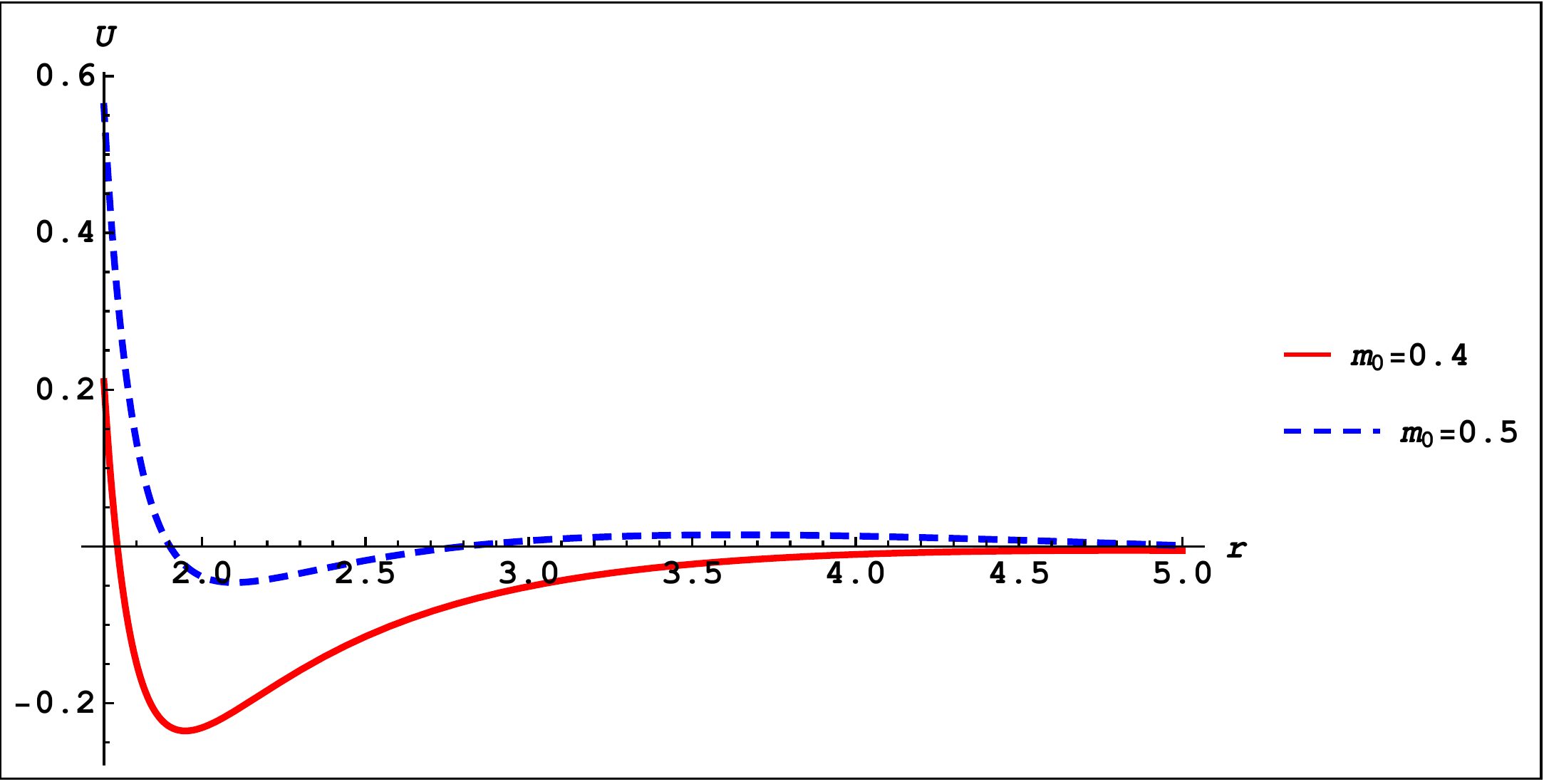}
	\includegraphics[width=8cm,height=5.5cm]{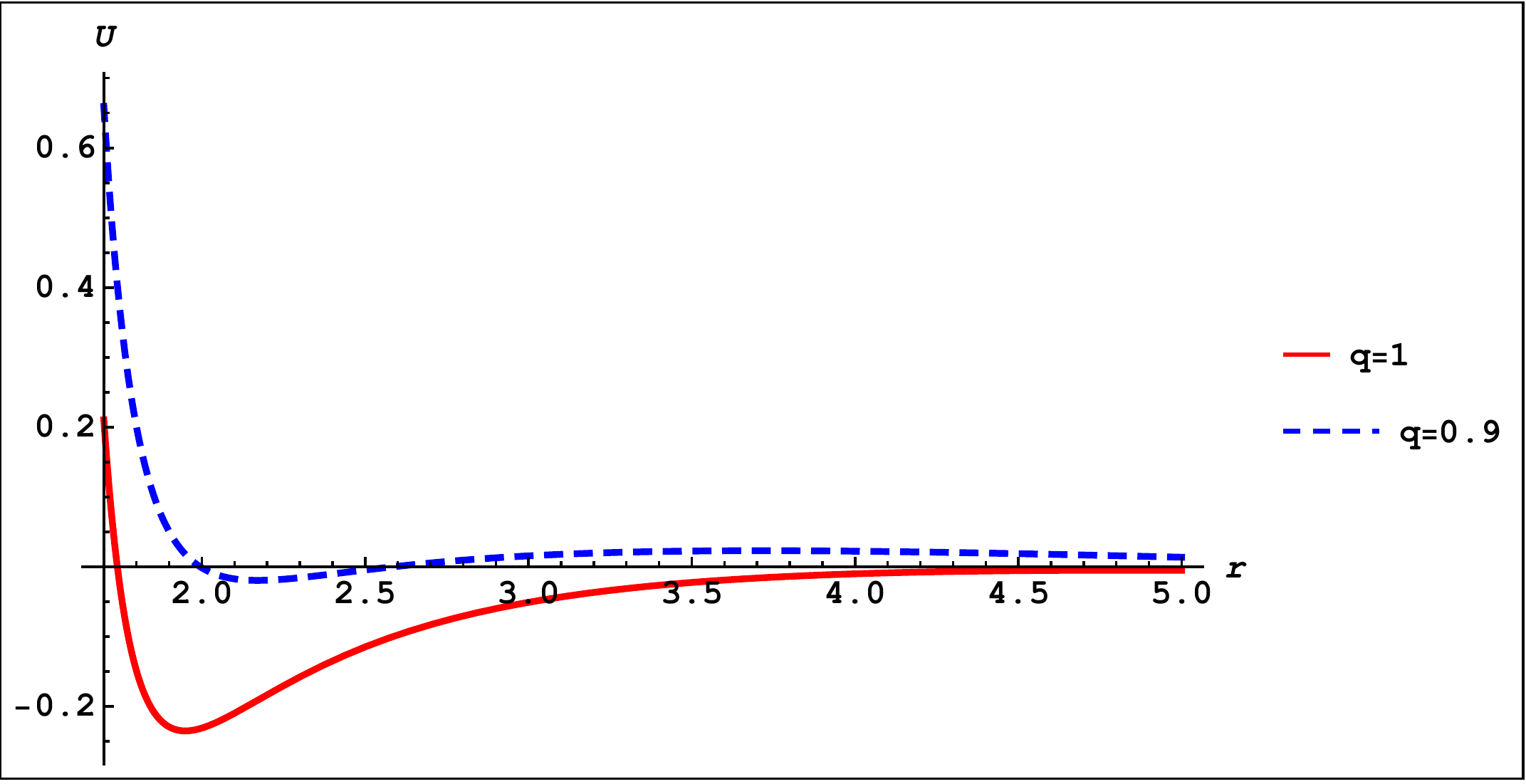}
	\caption{The shapes of the radial Schrodinger potential $U$ as a function of $r$ for model (\ref{model}) with $\alpha=0.001$, $\beta=3$, $n=1$, $L=10$, $Q=0.9$, $M=1$, Left: $q=1$ and different values of the scalar mass. Right: $m_0=0.4$ and different values of the scalar charge.  }
	\label{welll}
\end{figure}

In order to trigger superradiant instability, the following two conditions need to be satisfied simultaneously; trapped modes amplification must occur in a certain range of the frequency (superradiant regime) which is confirmed by Fig. \ref{fig3} and, a trapping potential well has to emerge outside the event horizon of the RN black hole to confine the scalar field as is clearly seen in Fig. \ref{well}. The plot shows that  scalar field will condense in the well and form the scalar hair. It is worth mentioning that to plot $U$ as a function of $r$, we  define $\Delta \equiv r^2 N$ and $R\equiv \frac{\sqrt{\Delta} \Psi}{r}$ \cite{sahar}. This enables equation (\ref{24}) to be written in the form of a Schrodinger-like wave equation $\frac{d^2 R}{dr^2}+U R=0$ where \cite{sahar, hor, zo, ar}
\begin{equation}U=\frac{r^4(\omega +q \phi)^2-\Delta \big(l(l+1)+r^2 m_0^2\big)-\frac{1}{2}\Delta \Delta'' +\frac{1}{4}{\Delta'}^2}{\Delta^2}.
\end{equation}
As can be seen in Fig. \ref{welll}, increasing the scalar charge or decreasing the scalar mass leads to the growth of depth of the potential which is compatible with the results presented in Fig. \ref{fig3}. The growing depth of the potential indicates amplification of superradiant instability.


In a process similar to planar hairy black holes, we seek static solutions with a nontrivial scalar field in the nonlinear regime by turning on the scalar field near critical frequency. Equations (\ref{14}) and (\ref{17}) now become
\begin{equation}\label{28}
\phi'(r)^2+m_{0}^2 h(r) \psi(r)^2=-\frac{2(1+f_{R})}{r}\left[h(r)\left(N'(r)+\frac{N(r)}{r}-\frac{1}{r}\right)+\frac{1}{2} N(r) h'(r)\right] -f_{R} h(r) R_{0}+f h(r),
\end{equation}
\begin{equation}\label{29}
N'_{h}=-\frac{r_{h}}{2(1+f_{R_{0}})}\left(\frac{{\phi'_{h}}^2}{h_{h}}+m_{0}^{2} \psi_{h}^2-f+f_{R_{0}}R_{0} \right)+\frac{1} {r_{h}},
\end{equation}
corresponding to metric (\ref{22}). We get the same boundary conditions as in section \ref{BHb} but as the symmetry of equations has changed \cite{bas}, we just fix $AdS$ boundary $L=1$ and leave $r_{h}$ for small black holes $\frac{r_{h}}{L}\ll1$ as a free parameter.
\subsection{ Numerical solution\label{BH1}}
In this section all the plots are drawn in a way akin to the method of  \ref{BH1a} where $L=1$ is used to re-scale the new set of equations. Throughout this section we use (\ref{model}) while the parameters remain unchanged and we use $q^{2}\geq \frac{9}{2}$ to turn off the near horizon instability \cite{oscar1} and only focus on superradiance in a global $AdS$ space-time.
\begin{figure}[!ht]
\includegraphics[width=8cm,height=5.5cm]{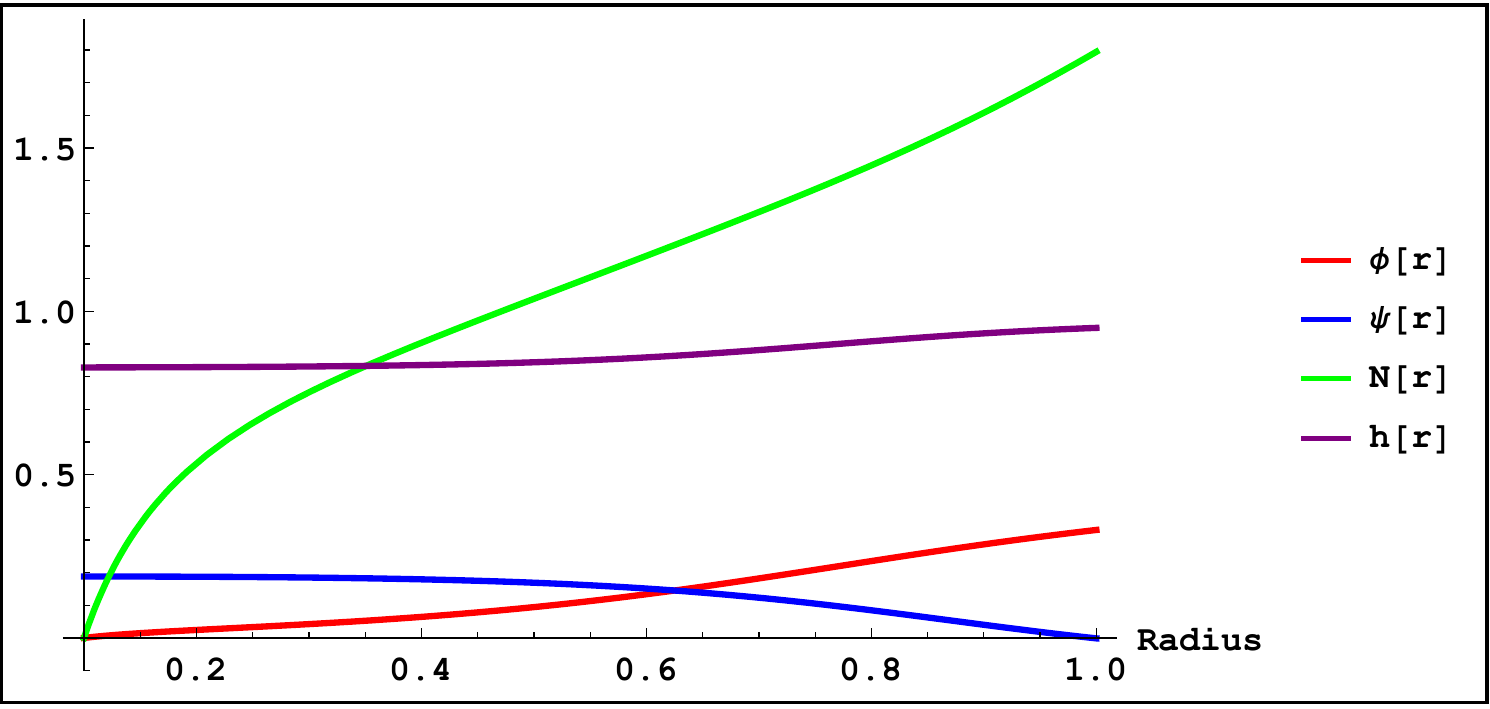}
\includegraphics[width=8cm,height=5.5cm]{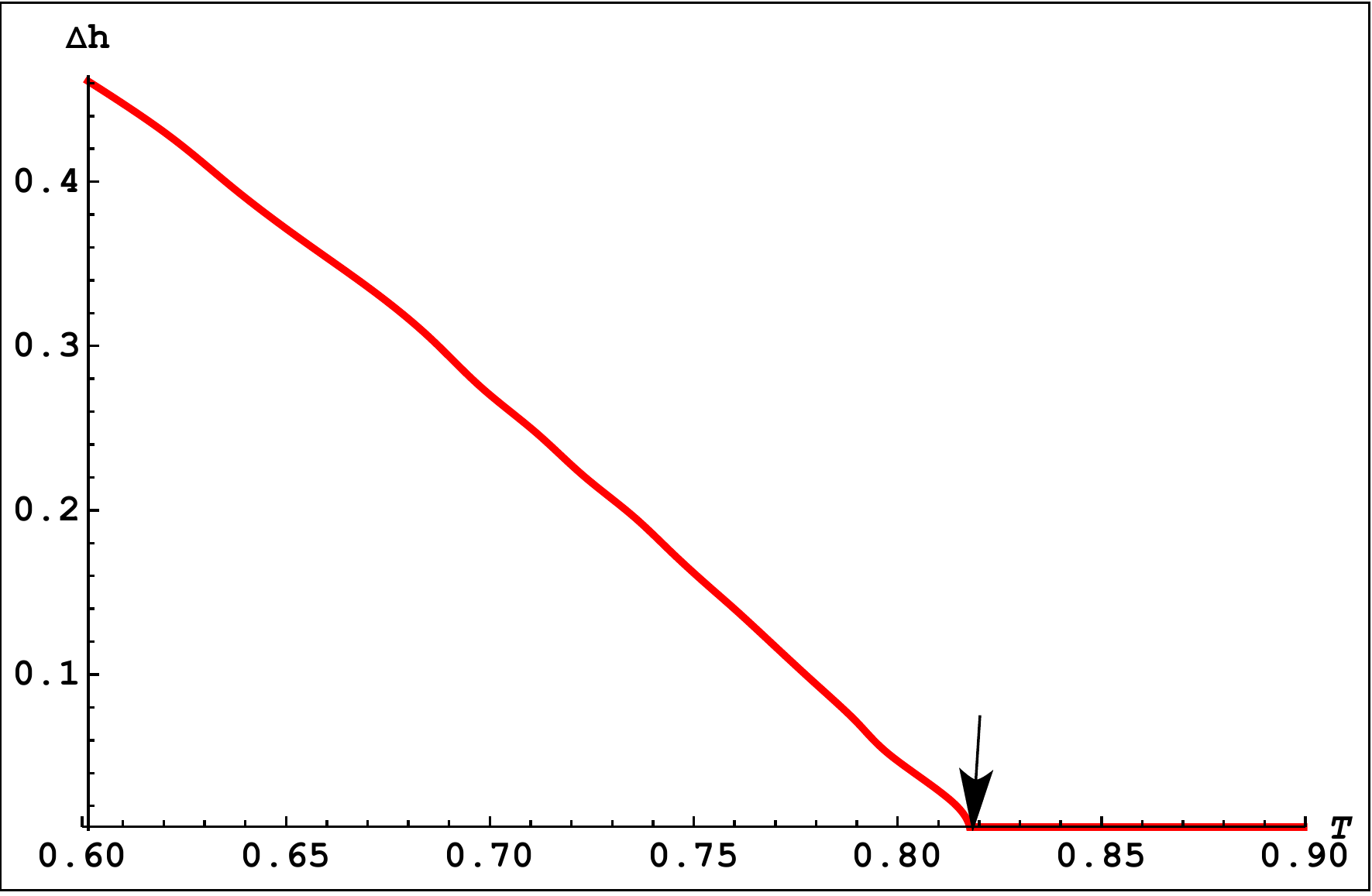}
\caption{Left: Plot of field variables as a function of radius with $q=30$, $m_{0}^{2}=0.04$, $r_{h}=0.1$, $\psi_{h}=0.188$ and $\phi'_{h}=0.4$. Right: Behavior of metric variable, $\Delta h$, as a function of temperature $T$ with $q=80$, $r_{h}=0.1$, $m_{0}^2=0$. The arrow shows critical temperature. }
\label{fig1s}
\end{figure}

\begin{figure}[!ht]
\includegraphics[width=8cm,height=5.5cm]{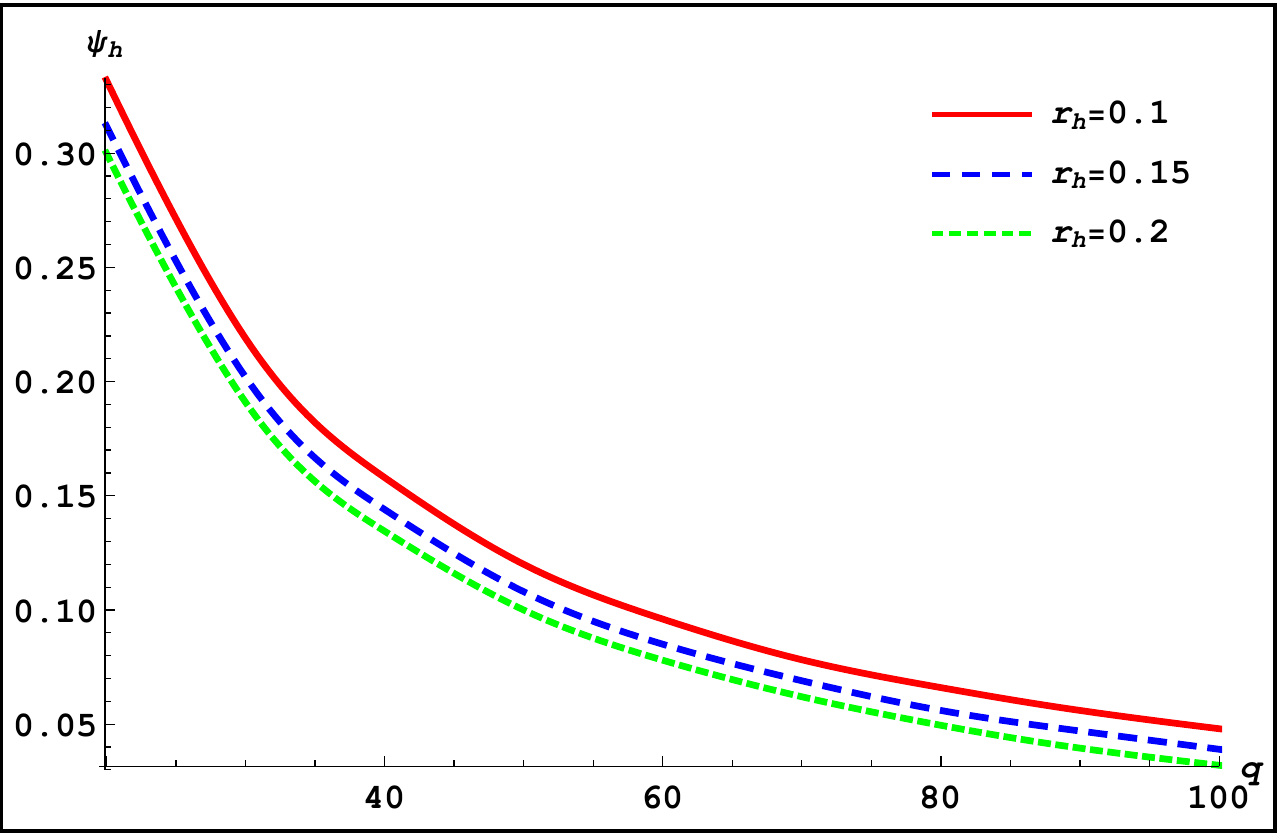}
\includegraphics[width=8cm,height=5.5cm]{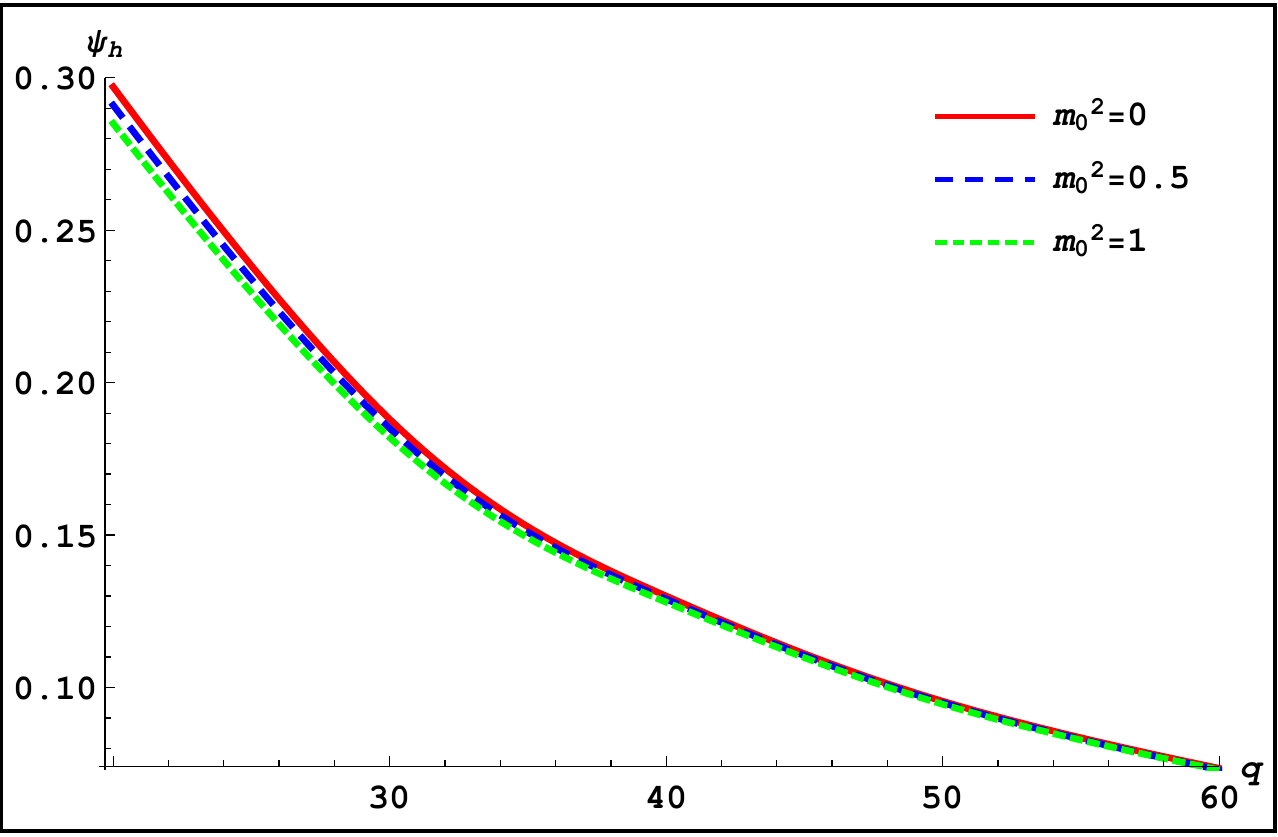}
\caption{$\psi_{h}$ is plotted as a function of $q$ when the scalar field has only one node at the reflective boundary. Left:  $\phi'_{h}=0.2$, $m_{0}^{2}=0$ and different horizon radii. Right:  $\phi'_{h}=0.4$, $r_{h}=0.1$ and different values of $m_{0}^2$. }
\label{fig2s}
\end{figure}

The left panel in Fig. \ref{fig1s} implies the existence of regular and nonsingular solutions outside the horizon. As can be seen, the oscillatory profile of the scalar field and other solutions depends on initial conditions.  The behavior of metric variable $h(r)$ demonstrates back reaction of the scalar field on space-time geometry. In the right panel, the change of metric variable $\Delta h=h_{\infty}-h_{h}$ as a function of $T$ is shown. According to the plot, the back reaction of the scalar field on background increases for $T<T_{c}$ while it almost vanishes for $T>T_{c}$, i.e. there is a hairy black hole for $T<T_{c}$ but for $T>T_{c}$, only $AdS$ RN black holes exist. Note that metric variable $h(r)$ also shows phase transition of RN black holes to hairy black holes at the critical temperature and the arrow shows critical temperature where phase transition occurs. The critical temperature is the point at which $\psi_{2}$ condensation  vanishes. Note that the asymptotic behaviour of the vector potential and scalar field are assumed as that given by equation (\ref{19}). The free parameters $q$, $\psi_{h}$, $\phi'_{h}$ and $r_{h}$  show phase space of the solutions. Fig. \ref{fig2s} shows dependence of $\psi_{h}$ as a function of $q$ for different values of  horizon radii and $m_{0}^2$ and as can be seen, the smaller the horizon radius, the larger the value of $\psi_{h}$, resulting in more freedom for choosing $\psi_{h}$. The right panel in Fig. \ref{fig2s} demonstrates that $\psi_{h}$ has an inverse relation to the scalar mass for small scalar charges. However, the scalar mass has no effect on $\psi_{h}$ for large scalar charges.

\begin{figure}[!ht]
\includegraphics[width=8cm,height=5.5cm]{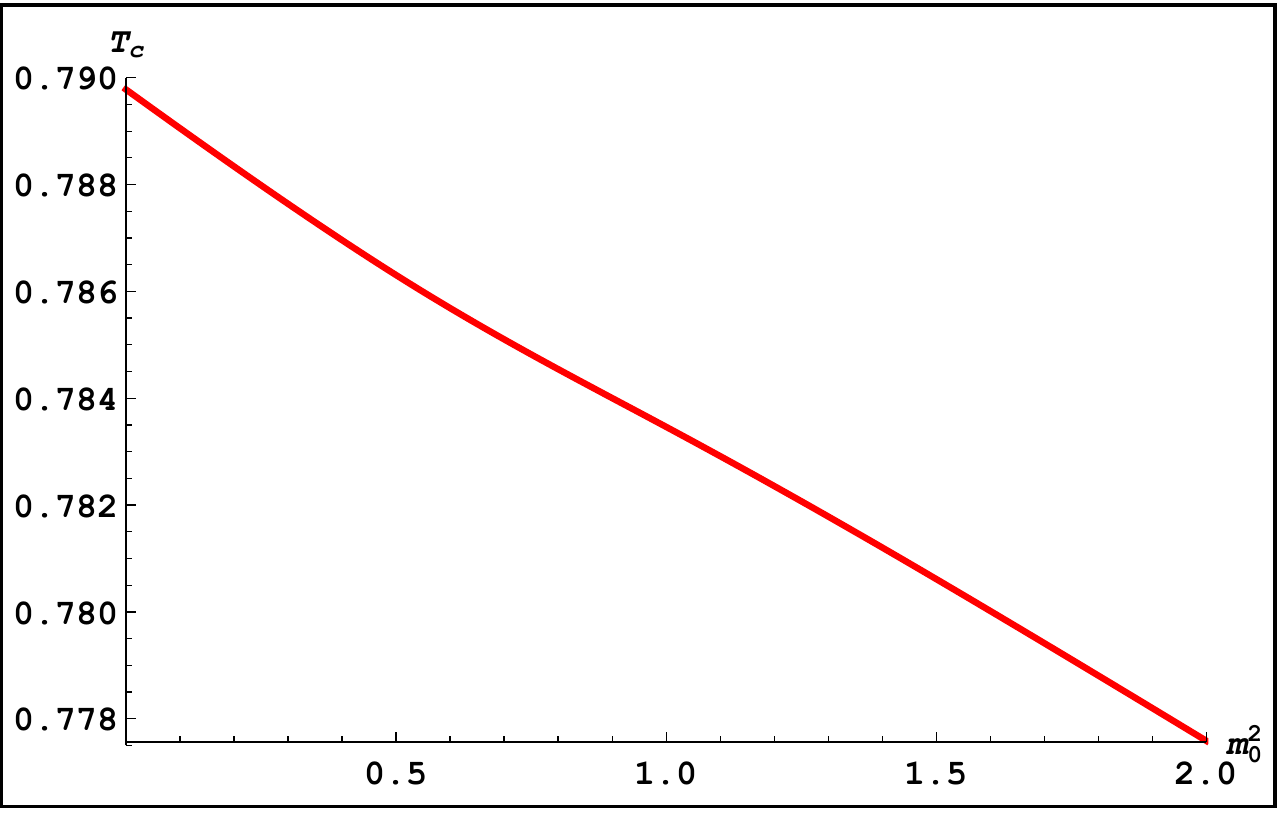}
\includegraphics[width=8cm,height=5.5cm]{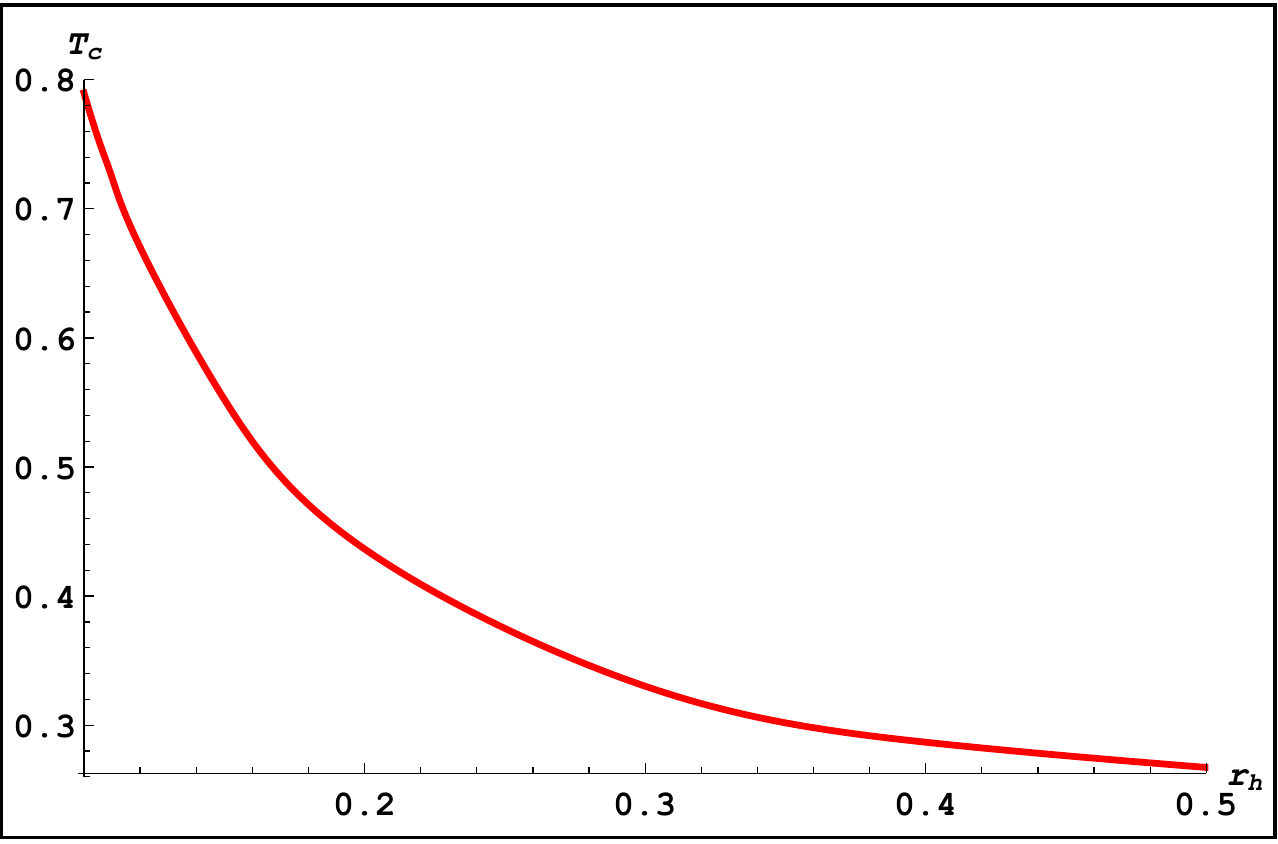}
\caption{The critical temperature for $q=40$ as a function of, Left: $m_{0}^2$  and $r_{h}=0.1$. Right: $r_{h}$ and  $m_{0}^2=0$. }
\label{fig3s}
\end{figure}

\begin{figure}[!ht]
\includegraphics[width=8cm,height=5.5cm]{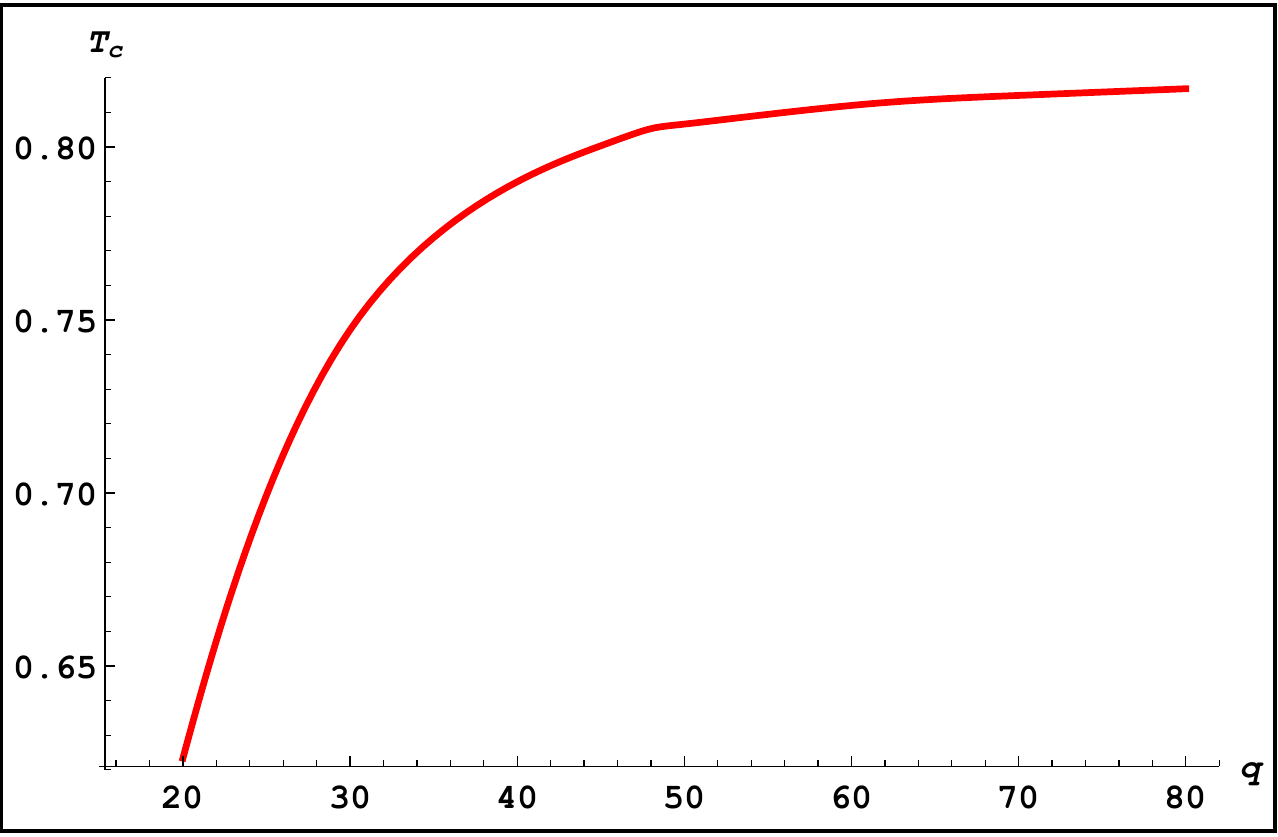}
\includegraphics[width=8cm,height=5.5cm]{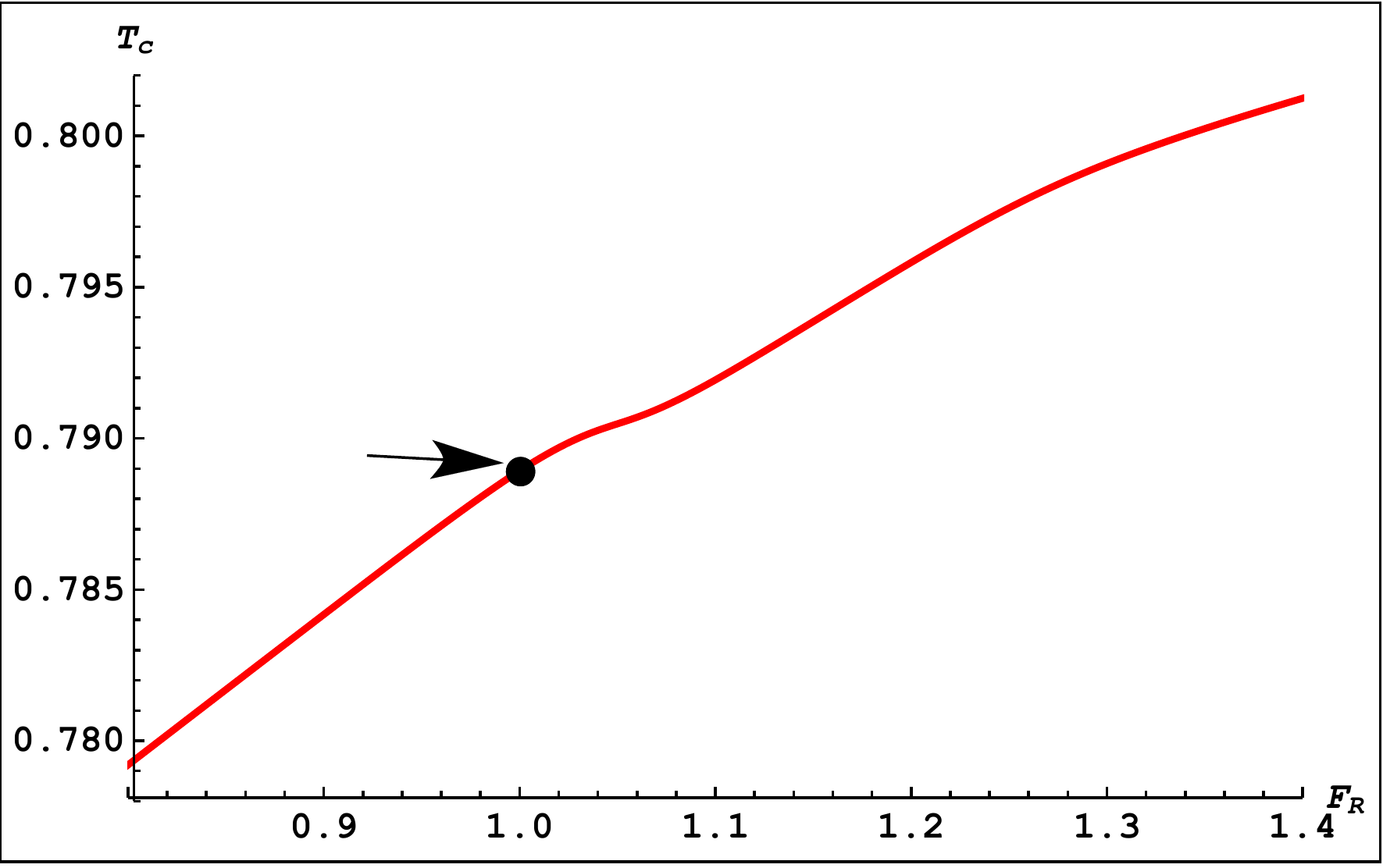}
\caption{Left:  Critical temperature as a function of scalar charge with $m_{0}^2=0$ and $r_{h}=0.1$. Right:  $F_{R}$ as a function of critical temperature for $q=40$, $r_{h}=0.1$ and $m_{0}^2=0$. The arrow shows the critical temperature in $AdS$ Einstein-Maxwell scalar field theory. }
\label{fig5s}
\end{figure}
Fig. \ref{fig3s} shows that critical temperature has an inverse relation to the scalar mass and event horizon. As can be seen in the left panel of Fig. \ref{fig3s}, for small values of scalar mass we see phase transition at high temperature, that is,  phase transition occurs earlier. In the right panel, by increasing size of the black hole ($0.3<r_{h}<0.5$), critical temperature decreases very slowly.
The left panel in Fig. \ref{fig5s} also shows relation between critical temperature and scalar charge which first increases by increasing the scalar charge, but asymptotically reaches a constant value. Finally the right panel in Fig. \ref{fig5s} shows the effect of $F_{R}$ on the possibility of  phase transition. It can readily be seen that as $F_{R}$ increase, there is a higher chance for the system to experience phase transition for higher critical temperatures. The arrow indicates the point at which phase transition occurs.

\begin{figure}[!ht]
	\includegraphics[width=10cm,height=5.5cm]{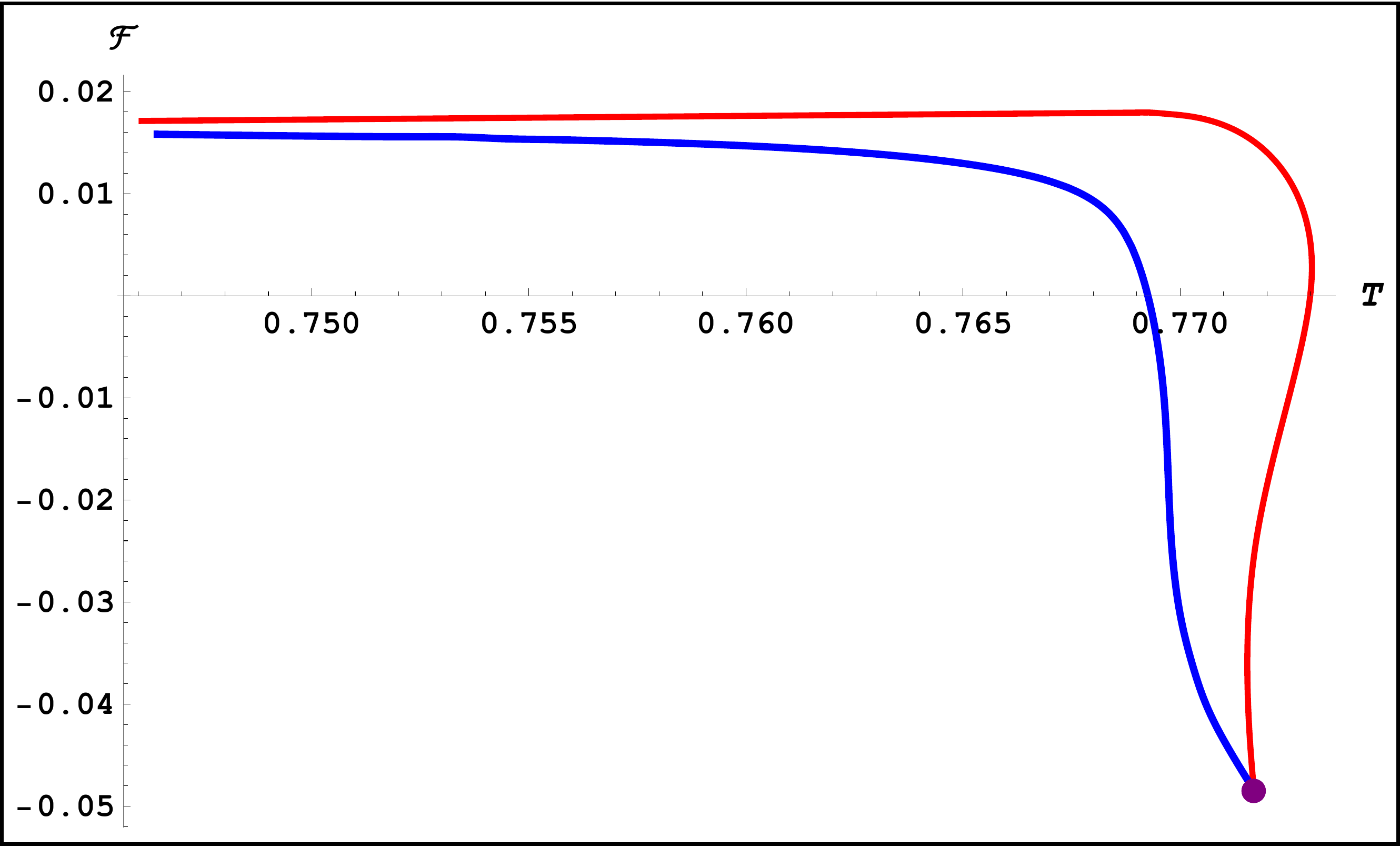}
	\caption{Free energy $\cal F$ as a function of  temperature $T$ for $r_h=0.1$, $m_0=0$ and $q=40$. The red line shows free energy for a small $AdS$ RN black hole and blue line is for a hairy black hole. The solid point represents the critical transition phase. }
	\label{free}
\end{figure}
To gain a deeper understanding of phase transition properties and thermodynamic stability of black holes, studying the behaviour of free energy would be in order at this point. The free energy $\cal F$ is given by
\begin{equation}\label{29d}
{\cal F}=E-T S -\mu Q,
\end{equation}		
where $S$ is the entropy and $E$ the energy of the system. As is shown in \cite{del}, thermodynamic properties of $F(R)$ black holes with constant negative curvature are qualitatively similar to those of $AdS$ black holes\footnote{In fact for $F(R)=R+f(R)$ models, the effective constant is obtained in term of $f(R)$.}.  One may evaluate the free energy of a gravitational system by evaluating the classical action directly, $S_e=\beta {\cal F}$, where $S_e$ is the classical action and $\beta=\frac{1}{T}$. However, the classical energy evaluated by this method is divergent. To make it finite a new action, $S_{reg}$, was introduced in \cite{bas}   by subtracting out the classical action of the global $AdS$ solution, leading to  a well-defined free energy given by ${\cal F}=\frac{S_{reg}}{\beta} $. In \cite{bas}, it was shown that both approaches, that is ${\cal F}=\frac{S_{reg}}{\beta} $ and equation (\ref{29d}), lead to a similar result for free energy for RN black holes. Following \cite{bas}, we plot the free energy $\cal F$ as a function of temperature $T$ by changing the initial condition in Fig. \ref{free} and compare the free energy for the hairy black hole phase and small RN black holes at the same interval ($T<T_c$) \cite{bas, bas1}. As can be seen in Fig. \ref{free}, the blue line represents the free energy of a hairy black hole while the red line refers to a small $AdS$ RN black hole and the solid point shows the critical transition phase. It is seen that the free energy of a hairy black hole configuration for $T<T_c$ is smaller than that for a small $AdS$ RN black hole which indicates that the hairy solution is the thermodynamically dominant and favoured phase.

\subsection{stability analysis\label{BHH1}}
Our results in the previous subsection is based on transition of the system from asymptotically RN-{$AdS_{4}$} as the background to a stable hairy configuration because of superradiant instability. To confirm the existence of hairy solution derived in \ref{BH1}, we have to  study the time evolution of the system. To do so, we assume that all field variables, in addition to radial dependence, are time-dependent. By defining $\xi=N \sqrt{h}$ and manipulating  equations (\ref{4})-(\ref{6}), we obtain the following dynamical equations
\begin{eqnarray}\label{30}
&&\frac{(1+f_{R})N'}{ N}+\frac{f_{R} R r-r f(R)}{2 N}-\frac{(1+f_{R})(1-N)}{N r}=-\frac{r}{2\xi^2}\left(|\dot{\psi}|^2+|\xi \psi'|^2+q^2|\phi|^2|\psi|^2+2q\phi\mbox{Im}(\psi\dot{\psi}^*)\right.\nonumber \\
&&\left.+N\phi'^2+m_{0}^2N h \psi^{2}\right),
\end{eqnarray}
\begin{eqnarray}\label{31}
\frac{(1+f_{R})h'}{h}=\frac{r}{\xi^2}\left(|\dot{\psi}|^2+|\xi \psi'|^2+q^2 |\phi|^2|\psi|^2+2q\phi\mbox{Im}(\psi\dot{\psi}^*)\right),
\end{eqnarray}
\begin{equation}\label{32}
 \frac{(1+f_{R}) \xi'}{\xi}+\frac{f_{R}R r}{2 N}-\frac{r f(R)}{2 N} =-\frac{r N \phi'^2}{2\xi^2}-\frac{r m_{0}^2 \psi^{2}}{2N}+\frac{(1+f_{R})(1-N)}{Nr} ,
\end{equation}
\begin{equation}\label{33}
-\frac{(1+f_{R})\dot{N}}{N}=r \mbox{Re}(\dot{\psi}^*\psi')+r q \phi\mbox{Im}(\psi'^*\psi).
\end{equation}
From Maxwell equation (\ref{5}), we obtain two dynamical equations
\begin{equation}\label{34}
\frac{\xi}{r^2}\left(\frac{r^2\phi'}{h^{\frac{1}{2}}}\right)'=q^2|\psi|^2 \phi-q\mbox{Im}(\dot{\psi}\psi^*),
\end{equation}
\begin{equation}\label{35}
\frac{1}{r}\partial_{t}{\left(\frac{r\phi'}{h^{\frac{1}{2}}}\right)}=-q\mbox{Im}(\xi \psi'\psi^*).
\end{equation}
Now, defining $\psi=\frac{\Psi}{r}$, the Klein-Gordon equation (\ref{6}) is given by
\begin{equation}\label{36}
-\ddot{\Psi}+\left(\frac{\dot{\xi}}{\xi}+2iq\phi\right)\dot{\Psi}+\xi(\xi \Psi')'+\left(iq\dot{\phi}-\frac{\xi \xi'}{r}-iq\frac{\dot{\xi}}{\xi}\phi+q^2\phi^2-\frac{\xi^2}{N} m_{0}^2\right)\Psi=0.
\end{equation}
where a dot signifies partial derivative with respect to $t$.
To investigate stability of static solutions, we consider linear perturbations around such solutions and define $N(r,t)=\bar{N}+\delta N(r,t)$ and similarly for other variables, where $\bar{N}$ are the static solution and $\delta N(r,t)$ shows perturbation part. By substituting linear perturbations in equations (\ref{30}-\ref{36}) and defining $\delta \Psi =\delta u+i \delta \dot{w}$,  one finds three perturbation equations for $\delta u$, $\delta w$ and $\delta \phi$ which constitute two dynamical equations and a constraint, see Appendix A for details.

\begin{figure}[!ht]
\includegraphics[width=8cm,height=5.5cm]{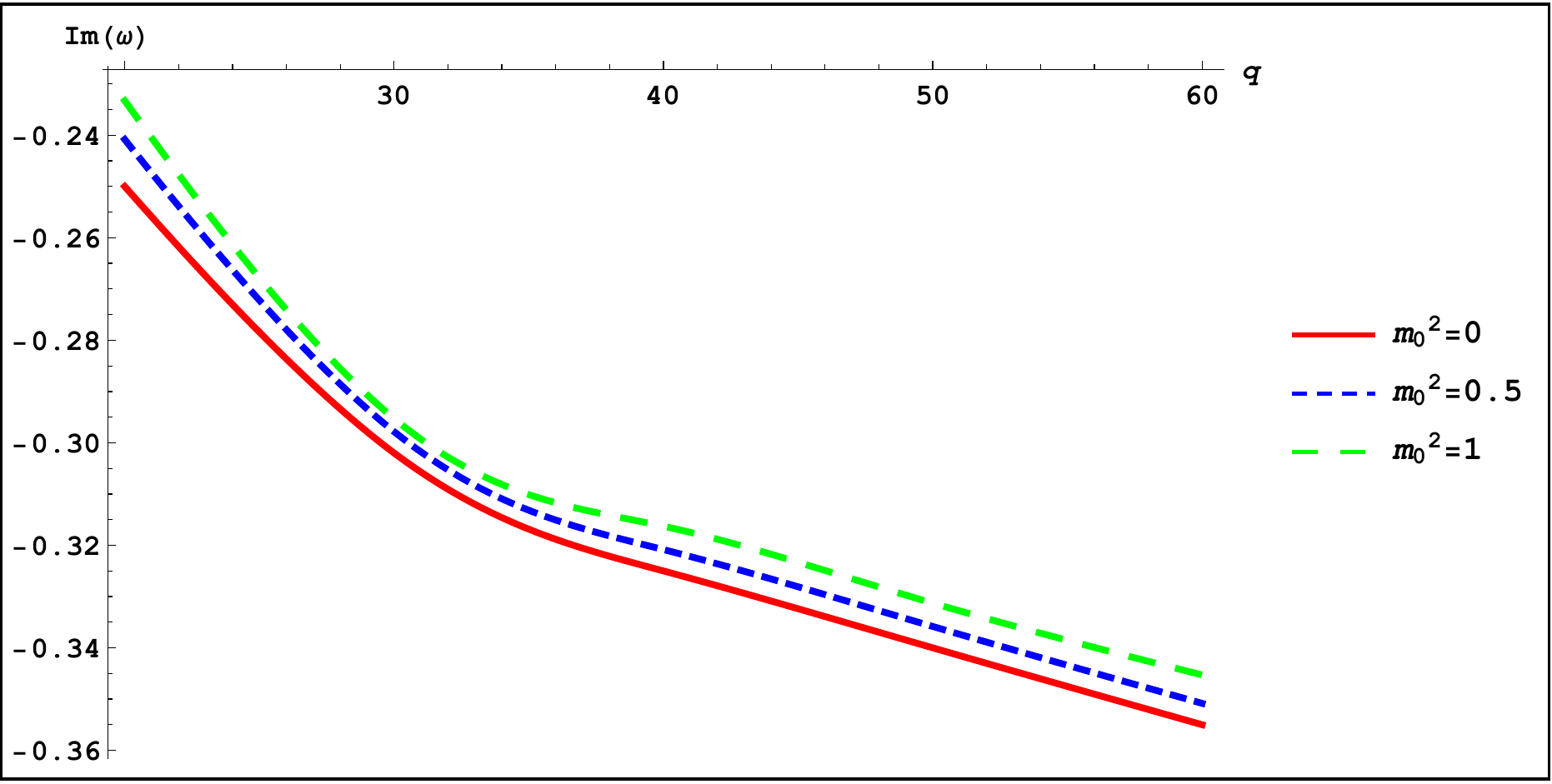}
\includegraphics[width=8cm,height=5.5cm]{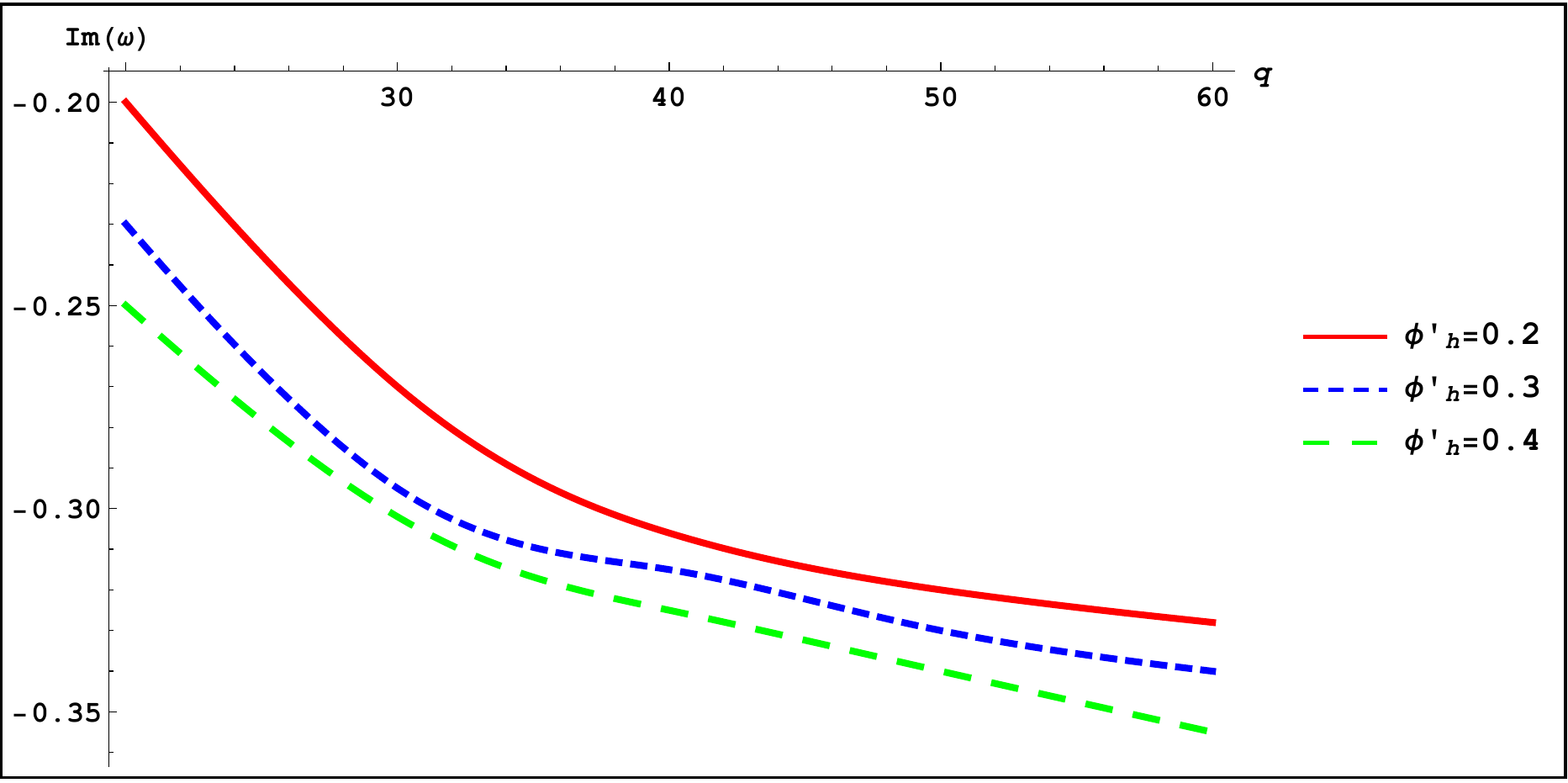}
\caption{\footnotesize{Imaginary parts of perturbation modes plotted as a function of $q$ for $r_{h}=0.1$,  Left: $\phi'_{h}=0.2$ and different values of scalar mass. Right: $m_{0}^{2}=0$ and different values of $\phi'_{h}$, when the static solutions of the scalar field have only one node at the reflective boundary for the selected model. Here, the perturbation modes decay exponentially in time since the imaginary part of the frequency is negative.}}\label{fig9}
\end{figure}

Due to linearity of perturbed equations and boundary conditions, we integrate coupled perturbed equations (\ref{50}-\ref{52}) using the shooting method and boundary conditions (\ref{305a}). Shooting parameters $\tilde{u}_{0}$ and $\omega$ are determined in such a way as to make scalar modes to vanish at the reflective boundary. In Fig. \ref{fig9},  using the phase space of solutions presented in Fig. \ref{fig2s}, we have plotted the imaginary part as function of $q$ for different values of the scalar mass and $\phi'_h$. As can be seen, the sign of $\mbox{Im}(\omega)$ is negative for one node solutions at the critical temperature which represents perturbation modes decaying exponentially and the hairy black hole becoming stable. Thus hairy solutions at the critical temperature are stable and one may consider those as a possible endpoint of superradiant instability.
\section{Discussion and conclusions }\label{DC}
In this paper we have studied instabilities of $F(R)$-charged $AdS_{4}$ black holes caused by a massive charged scalar field. We investigated the near horizon scalar condensation instability in a certain model of modified gravity for a charged black hole with planar horizon and show that this type of instability finally leads to a hairy black hole configuration. We numerically verified the existence of hairy black hole solutions and presented the phase space of solutions and showed that such a transition is of second order and condensation exists for $T<T_{c}$. Also, an interesting result was presented in the right panel of Fig. \ref{fig2p} which shows that for higher values of $F_{R}$, phase transition occurs at higher critical temperatures, leading to a new generation of high temperature superconductors in the context of $F(R)$ gravity.

Small global black holes in $F(R)$-charged scalar field theory suffer from superradiant instability in a certain range of frequencies associated with scalar field wave function. We found that $\mbox{Im}(\omega)$ becomes larger by increasing $q$ and decreasing $m_{0}$. Besides, there is no superradiant instability for $q=m_{0}$ or $q<m_{0}$. We showed numerically that black hole solutions with charged scalar hair are admitted in $F(R)$-charged $AdS_{4}$ scalar field theory. Also,  the solution space becomes smaller by increasing the scalar mass, scalar charge and the event horizon.
Since the behaviour of metric variable $h(r)$ is an indication of phase transition of a RN black hole to a hairy solution, the right panel in Fig. \ref{fig1s}  shows that the system settles in a RN black hole configuration for $T>T_{c}$, while a hairy black hole configuration is preferred for $T<T_{c}$ where the phase transition occurs at $T_{c}$.
We also investigated dependence of $T_{c}$ on scalar mass, scalar charge and event horizon. We showed that $T_{c}$ has an inverse relation to the scalar mass and event horizon unless   $0.3<r_{h}<0.5$, for which the critical temperature decreases slowly with the growing size of the black hole.
In addition, we showed that for larger values of $F_{R}$, the system will experience  higher critical temperatures and phase transition becomes more likely.  On the other hand, it was shown that the one node scalar field profiles at critical temperature which is connected to the onset of superradiance, are stable. Such hairy black holes may then be considered as natural candidates for the endpoint of superradiant instability.

\section{Acknowledgments}
M. Honardoost would like to thank Iran National Science Foundation (INSF) and Research Council of Shahid Beheshti University for financial support.
\vspace{3mm}
 \appendix
 \section{ Appendix A}
\subsection{Perturbation equations }
We define the perturbed quantities according to $N(t,r)=\bar{N}+\delta N(t,r)$ and in a similar manner for other quantities where $\bar{N}$ is the equilibrium value obtained from static equations and $\delta N$ shows perturbation part. One then obtains perturbations of dynamical equations (\ref{30}-\ref{36}) to first order 
\begin{eqnarray}\label{38}
&&\frac{(1+\bar{f}_{R})\delta N'}{\bar{N}}-\left[\frac{1+\bar{f}_{R}}{\bar{N}}\left(\bar{N'}-\frac{1}{r}\right)-\frac{r \bar{\phi'}^2}{2 \bar{N}\bar{h}}-\frac{\bar{f}r}{2\bar{N}}+\frac{\bar{f}_{R}r\bar{R}}{2\bar{N}}+\frac{ m_0^2 \bar{\Psi}^2}{2 r \bar{N}}\right]\frac{\delta N}{\bar{N}}=\left(\frac{q^2 \bar{\phi}^2 \bar{\Psi}^2}{r \bar{\xi}^2}+\frac{r \bar{\phi'}^2}{\bar{N}\bar{h}}\right)\frac{\delta \xi}{\bar{\xi}}\nonumber \\
&&-\frac{q^2 \bar{\phi} \bar{\Psi}^2}{r \bar{\xi}^2}\delta \phi-\frac{r \bar{\phi'}}{\bar{N}\bar{h}} \delta \phi'-\bar{f}_{RR} \left(\frac{1}{\bar{N}r}(\bar{N}-1)+\frac{ r \bar{R_{0}}}{2 \bar{N}}+\frac{\bar{N}'}{\bar{N}}\right) \delta R-\frac{i q \bar{\phi} \bar{\Psi}}{2 r \bar{\xi}^2}(\delta \dot{\Psi}-\delta \dot{\Psi}^\ast) -\left(\frac{\bar{\Psi}}{2 r}\right)' (\delta \Psi'+\delta \Psi'^\ast)\nonumber \\
&&+\left[\frac{\left(\frac{\bar{\Psi}}{r}\right)'}{2 r}-\frac{q^2\bar{\phi}^2 \bar{\Psi}}{2 r \bar{\xi}^2}-\frac{m_0^2 \bar{\Psi}}{2 r \bar{N}}\right](\delta \Psi+\delta \Psi^\ast),
\end{eqnarray}
\begin{eqnarray}\label{39}
&&\frac{(1+\bar{f}_{R})\delta h'}{\bar{h}}-\frac{(1+\bar{f}_{R})\bar{h'}\delta h}{\bar{h}^2}+\frac{\bar{f}_{RR}\bar{h'}\delta R}{\bar{h}}=\frac{i q \bar{\phi} \bar{\Psi}}{r \bar{\xi}^2}(\delta \dot{\Psi}-\delta \dot{\Psi}^\ast) -\left[\frac{1}{r}\left(\frac{\bar{\Psi}}{r}\right)'-\frac{q^2 \bar{\phi}^2 \bar{\Psi}}{r \bar{\xi}^2}\right](\delta \Psi+\delta \Psi^\ast) \nonumber \\
&&+\left(\frac{\bar{\Psi}}{r}\right)'(\delta \Psi'+\delta \Psi'^\ast)+\frac{2 q^2 \bar{\phi} \bar{\Psi}^2}{r \bar{\xi}^2} \delta \phi-\frac{2 q^2 \bar{\phi}^2 \bar{\Psi}^2}{r \bar{\xi}^3}\delta \xi,
\end{eqnarray}
\begin{eqnarray}\label{40}
&&\frac{(1+\bar{f}_{R})\delta \xi'}{\bar{\xi}}=-\left[\frac{r\bar{\phi'}^2}{2 \bar{h}}+\frac{1}{r}(1+\bar{f_{R}})+\frac{\bar{f} r}{2 }-\frac{\bar{f}_{R} \bar{R_{0}}r}{2}-\frac{m_0^2\bar{\Psi}^2}{2 r}\right]\frac{\delta N}{\bar{N}^2}+\left[\frac{(1+\bar{f}_{R})\bar{\xi}'}{\bar{\xi}}+\frac{r\bar{\phi'}^2}{\bar{N} \bar{h}}\right]\frac{\delta \xi}{\bar{\xi}}-\left(\frac{1}{r}\right.\nonumber \\
&&\left.-\frac{1}{\bar{N}r}+\frac{r\bar{R}}{2 \bar{N}}+\frac{\xi'}{\xi}\right)\bar{f}_{RR} \delta R-\frac{r \bar{\phi'}}{\bar{N}\bar{h}} \delta \phi'-\frac{m_0^2\bar{\Psi}}{2 r\bar{N}}(\delta \Psi+\delta \Psi^\ast),
\end{eqnarray}
where $\bar{f}_{RR}$ signifies $\frac{d^2\bar{f}}{dR^2}$. Equation (\ref{40}) is obtained by combining equations (\ref{38}) and (\ref{39}). The rest are given by
\begin{eqnarray}\label{41}
\frac{(1+\bar{f}_{R})\delta \dot{N}}{\bar{N}}=-\left(\frac{\bar{\Psi}}{2 r}\right)'(\delta \dot{\Psi}+\delta \dot{\Psi^\ast})+\frac{i q \bar{\phi} \bar{\Psi}'}{2r}(\delta \Psi-\delta \Psi^\ast)-\frac{i q \bar{\phi} \bar{\Psi}}{2r}(\delta \Psi'-\delta \Psi'^\ast),
\end{eqnarray}
\begin{eqnarray}\label{42}
&&\bar{N} \delta \phi''+\bar{N}\left(\frac{2}{r}-\frac{\bar{h}'}{2\bar{h}}\right) \delta \phi'-\frac{q^2 \bar{\Psi}^2}{r^2}\delta \phi=-\frac{q^2 \bar{\phi} \bar{\Psi}^2}{r^2 \bar{\xi}}\delta \xi+\frac{\bar{N}\bar{\phi'}}{2\bar{h}}\delta h'+\frac{1}{2\bar{h}}\left[\frac{q^2\bar{\phi}\bar{\Psi}^2}{r^2}-\frac{\bar{N}\bar{\phi'}\bar{h'}}{\bar{h}}\right]\delta h+\nonumber \\
&&\frac{i q  \bar{\Psi}}{2r^2}(\delta \dot{\Psi}-\delta \dot{\Psi^\ast})+\frac{q^2 \bar{\phi} \bar{\Psi}}{r^2}(\delta \Psi+\delta \Psi^\ast)
\end{eqnarray}
\begin{eqnarray}\label{43}
\frac{\delta \dot{ \phi'}}{\sqrt{\bar{h}}}-\frac{\bar{\phi'}\delta \dot{ h}}{2 \bar{h}\sqrt{\bar{h}}}=\frac{i q  \bar{\Psi} \bar{\xi}}{2 r^2}(\delta{\Psi'}-\delta{\Psi'^\ast})-\frac{i q \bar{\xi} \bar{\Psi}'}{2r^2}(\delta \Psi-\delta \Psi^\ast),
\end{eqnarray}
\begin{eqnarray}\label{44}
&&-\delta \ddot{\Psi}+\bar{\xi}^2 \delta \Psi''+2iq \bar{\phi} \delta \dot{\Psi}+ \bar{\xi} \bar{\xi}' \delta \Psi'+\left(q^2 \bar{\phi}^2-\frac{\bar{\xi}\bar{\xi}'}{r}-\frac{m_0^2 \bar{\xi}^2}{\bar{N}}\right)\delta \Psi-\frac{iq \bar{\phi} \bar{\Psi}}{\bar{\xi}} \delta \dot{\xi}+\left(\bar{\xi} \bar{\Psi}'-\frac{\bar{\xi} \bar{\Psi}}{r}\right)\delta \xi'+iq \bar{\Psi} \delta \dot{\phi}\nonumber\\
&&+\left(2 \bar{\xi} \bar{\Psi}''+\bar{\xi}'\bar{\Psi}'-\frac{\bar{\Psi} \bar{\xi}'}{r}-\frac{2\bar{\xi} m_0^2 \bar{\Psi}}{\bar{N}}\right)\delta \xi+2 q^2 \bar{\Psi}\bar{\phi} \delta \phi+\frac{m_0^2 \bar{\xi}^2\bar{\Psi}}{\bar{N}}\frac{\delta N}{\bar{N}}=0.
\end{eqnarray}
Since the scalar field is a complex quantity, we perturb it according to $\delta \Psi=\delta u+i \delta \dot{w}$. For the real part, we find
\begin{eqnarray}\label{45}
&&-\delta \ddot{u}+\bar{\xi}^2\delta u''-2q \bar{\phi}\delta \ddot{w}+\bar{\xi}\bar{\xi}'\delta u'+\left(q^2\bar{\phi}^2-\frac{\bar{\xi}\bar{\xi}'}{r}-\frac{m_0^2 \bar{\xi}^2}{\bar{N}}\right)\delta u+\left(\bar{\xi}\bar{\Psi}'-\frac{\bar{\xi}\bar{\Psi}}{r}\right)\delta \xi'+ \left(2\bar{\xi} \bar{\Psi}''+\bar{\xi}'\bar{\Psi}'-\frac{\bar{\Psi} \bar{\xi}'}{r}\right.\nonumber\\
&&\left.-\frac{2\bar{\xi} m_0^2 \bar{\Psi}}{\bar{N}}\right)\times \delta \xi+2q^2\bar{\Psi}\bar{\phi} \delta \phi+\frac{m_0^2 \bar{\xi}^2\bar{\Psi}}{\bar{N}}\frac{\delta N}{\bar{N}}=0,
\end{eqnarray}
and the imaginary part becomes
\begin{eqnarray}\label{46}
-\delta \dddot{w}+\bar{\xi}^2 \delta \dot{w}''+2q \bar{\phi} \delta \dot{u}+ \bar{\xi} \bar{\xi}' \delta \dot{w}'+\left(q^2 \bar{\phi}^2-\frac{\bar{\xi}\bar{\xi}'}{r}-\frac{m_0^2 \bar{\xi}^2}{\bar{N}}\right)\delta \dot{w}-\frac{q \bar{\phi} \bar{\Psi}}{\bar{\xi}} \delta \dot{\xi}+q \bar{\Psi} \delta \dot{\phi}=0.
\end{eqnarray}
Integration with respect to time of equations  (\ref{41}, \ref{43}) results in
\begin{eqnarray}\label{47}
(1+\bar{f_{R}})\frac{\delta N}{\bar{N}}=-\left(\frac{\bar{\Psi}}{r}\right)'\delta u-\frac{ q \bar{\phi} \bar{\Psi}'}{r}\delta w+\frac{q \bar{\phi} \bar{\Psi}}{r}\delta w'+\delta \emph{F(r)},
\end{eqnarray}
\begin{eqnarray}\label{48}
\frac{\delta h}{\bar{h} \sqrt{\bar{h}}}= \frac{2\delta  \phi'}{\bar{\phi'}\sqrt{\bar{h}}}+\frac{ 2 q  \bar{\Psi} \bar{\xi}}{r^2 \bar{\phi'}}\delta w'-\frac{ 2 q \bar{\xi} \bar{\Psi}'}{\bar{\phi'} r^2}\delta w+\delta \textit{g(r)},
\end{eqnarray}
where $\delta \emph{F(r)}$ and $\delta \emph{g(r)}$ are arbitrary functions of the $r$. Now integration of equation (\ref{46})  gives
\begin{eqnarray}\label{49}
\delta \ddot{w}-\bar{\xi}^2 \delta w''-2q \bar{\phi} \delta u-\bar{\xi} \bar{\xi}' \delta w'-\left(q^2 \bar{\phi}^2-\frac{\bar{\xi}\bar{\xi}'}{r}-\frac{m_0^2 \bar{\xi}^2}{\bar{N}}\right)\delta w+\frac{q \bar{\phi} \bar{\Psi}}{\bar{\xi}} \delta \xi-q \bar{\Psi} \delta \phi+\delta \textit{H(r)}=0,
\end{eqnarray}
where $\delta \textit{H(r)}$ appears as an arbitrary function of $r$. Due to the form of $\delta \Psi$, provided an arbitrary function of $r$ is added to $\delta w$,  it remains constant which allows us to set $\delta \textit{H(r)}=0$.

A first order differential equation is formed by plugging (\ref{38}) in (\ref{42}), using equations (\ref{47} -\ref{49})
\begin{eqnarray}\label{49a}
&&\delta \emph{F(r)}'+\left(\frac{\bar{\xi}'}{\bar{\xi}}+\frac{1}{r}\right)\delta \emph{F(r)}+\bar{f}_{RR}\left(\frac{1}{r}-\frac{1}{\bar{N}r}+\frac{r\bar{R}}{2 \bar{N}}+\frac{\bar{N}'}{\bar{N}}\right) \delta R=\frac{r \bar{\phi}\bar{\phi'}\bar{N}}{\bar{\xi}^2}(\sqrt{\bar{h}} \delta \emph{g(r)})'+\nonumber\\
&&\frac{r \bar{\phi'}^2}{2\bar{\xi}}\delta \emph{g(r)}+\frac{q^2\bar{\phi}^2\bar{\Psi}^2\sqrt{\bar{h}}}{2r\bar{\xi}^2}\delta \emph{g(r)}.
\end{eqnarray}
Now, substitution  of (\ref{38}) and (\ref{42}) in (\ref{47} ) leads to another first order differential equation  
\begin{equation}\label{49b}
\delta \emph{F(r)}'+\left(\frac{\bar{\xi}'}{\bar{\xi}}+\frac{1}{r}\right)\delta \emph{F(r)}+\bar{f}_{RR}\left(\frac{1}{r}-\frac{1}{\bar{N}r}+\frac{r\bar{R}}{2 \bar{N}}+\frac{\bar{N}'}{\bar{N}}\right) \delta R=\frac{r \bar{\phi'}^2}{2\bar{\xi}}\delta \emph{g(r)}.
\end{equation}
Since the ingoing boundary conditions for all perturbations, that is $\delta N$ and $\delta h$, must be satisfied \cite{win}, equations (\ref{47}) and (\ref{48}) must also satisfy such conditions, leading to $\delta \textit{F(r)}=0$ and $\delta \textit{g(r)}=0$ at $r=r_{h}$. So $\delta \textit{F(r)}$, $\delta \textit{g(r)}$, $\delta \textit{F(r)}'$ and $\delta \textit{g(r)}'$ are omitted from perturbation equations and $\delta R$ at $r=r_{h}$ needs to be zero due to equations (\ref{49a}, \ref{49b}).
Substitution of (\ref{47}) and (\ref{48}) in (\ref{49}) then the result in
\begin{eqnarray}\label{50}
&&\delta \ddot{w}-\bar{\xi}^2\delta w''+\left[\frac{q^2 \bar{\phi} \bar{\Psi}^2}{r^2  \bar{\phi}'}\left(\frac{r \bar{\phi}' \bar{\phi}}{1+\bar{f}_{R}}+ \bar{N}\bar{h}\right)-\bar{\xi} \bar{\xi}'\right] \delta w'-\left[\frac{q^2 \bar{\phi} \bar{\Psi} \bar{\Psi}'}{r^2  \bar{\phi}'} \left(\frac{ r \bar{\phi} \bar{\phi}'}{1+\bar{f}_{R}}+\bar{N}\bar{h}\right)+q^2 \bar{\phi}^2-\frac{\bar{\xi} \bar{\xi}'}{r}-m_{0}^ 2 \bar{N}\bar{h}\right] \nonumber \\
&&\times \delta w -q \bar{\phi}\left(2+\frac{\bar{\Psi} \left(\frac{\bar{\Psi}}{r}\right)'}{1+\bar{f}_{R}}\right)\delta u-q \bar{\Psi} \delta \phi+\frac{q \bar{\phi} \bar{\Psi}}{\bar{\phi}'} \delta \phi'=0.
\end{eqnarray}
Now, plugging equations (\ref{40}, \ref{47} -\ref{49}, \ref{49b}) in (\ref{45}), we find
\begin{eqnarray}\label{51}
&&\delta \ddot{u}-\bar{\xi}^2 \delta u''-\bar{\xi} \bar{\xi}'\delta u'+\left[3 q^2 \bar{\phi}^2+\frac{\bar{\xi} \bar{\xi}'}{r}+\frac{ \bar{N}\left(\frac{\bar{\Psi}}{r}\right)'^2 }{1+\bar{f}_{R}}\left(\frac{r^2 \bar{\phi}'^2}{2 (1+\bar{f}_{R})}-\frac{r^2 \bar{h} \bar{f}}{2( 1+\bar{f}_{R})}+\frac{r^2 \bar{h} \bar{f}_{R} \bar{R_{0}}}{2( 1+\bar{f}_{R})}-\bar{h}\right)+m_{0}^2\bar{N}\bar{h}\left(1+\right.\right.\nonumber \\
&&\left.\left.\frac{\bar{\Psi}(\frac{\bar{\Psi}}{r})'}{1+\bar{f}_{R}}\left(2+\frac{\bar{\Psi}(\frac{\bar{\Psi}}{r})'}{2\left(1+\bar{f}_{R}\right)}\right)\right)\right]\delta u+ 2 q \bar{\phi} \bar{\xi}^2 \delta w''+q \bar{N}\bar{\phi}\left[2 \sqrt{\bar{h}}\bar{\xi}'+\frac{\bar{\Psi} \left(\frac{\bar{\Psi}}{r}\right)'}{1+\bar{f}_{R}}\left(-\frac{\bar{N}\bar{h} \bar{\phi}' }{\bar{\phi}}-\frac{r \bar{\phi}'^2}{2(1+\bar{f}_{R})}-\frac{\bar{h}\bar{f}_{R} \bar{R_{0}} r}{2 (1+\bar{f}_{R})}+ \right.\right.\nonumber \\
&&\left.\left.\frac{\bar{h}}{r}+\frac{\bar{h}\bar{f} r}{2 (1+\bar{f}_{R})}\right)-\frac{m_{0}^2\bar{h}\bar{\Psi}^2}{r\left(1+\bar{f}_{R}\right)}\left(1+\frac{\bar{\Psi}(\frac{\bar{\Psi}}{r})'}{2\left(1+\bar{f}_{R}\right)}\right)\right]\delta w' +q \bar{\phi}\left[ 2 q^2 \bar{\phi}^2-\frac{2 \bar{\xi} \bar{\xi}'}{r}+\frac{\bar{\xi} \bar{\Psi}' \left(\frac{\bar{\Psi}}{r}\right)'}{1+\bar{f}_{R}}\left(\frac{\bar{\xi} \bar{\phi}'}{\bar{\phi}}-\bar{\xi}'-\frac{\bar{\xi}}{r }\right)+\right.\nonumber \\
&&\left.m_{0}^2\bar{N}\bar{h}\left(-2+\frac{\bar{\Psi}\bar{\Psi}'}{r\left(1+\bar{f}_{R}\right)}\right)\right]\delta w=0.
\end{eqnarray}
Equations (\ref{39}, \ref{48}, \ref{49}) then give
\begin{eqnarray}\label{52}
&&\frac{q \bar{\Psi}}{r}\left(\frac{\bar{\phi}}{1+\bar{f}_{R}}+\frac{\bar{N}\bar{h}}{r \bar{\phi}'}\right)\delta w''+ q \bar{\phi} \bar{\Psi} \left[-\frac{q^2 \bar{h} \bar{\Psi}^2}{r^4 \bar{\phi}'^2}+\frac{\bar{\xi}'}{(1+\bar{f}_{R}) \bar{\xi} r}+\frac{\sqrt{\bar{h}}\bar{\xi'}}{r^2 \bar{\phi} \bar{\phi}'}\right]\delta w'+\frac{q  \bar{\phi} \bar{\Psi}}{r^2}\left[\frac{r q^2 \bar{\phi}^2}{(1+\bar{f}_{R}) \bar{\xi}^2}+\frac{q^2 \bar{\phi}}{\bar{N} \bar{\phi}'}-\right.\nonumber\\
&&\left.\frac{\bar{\xi}'}{(1+\bar{f}_{R}) \bar{\xi}}-\frac{m_{0}^2 r}{\bar{N}(1+\bar{f_{R}})}-\frac{m_{0}^2\bar{h}}{\bar{\phi}\bar{\phi}'}-\frac{\sqrt{\bar{h}}\bar{\xi}'}{r \bar{\phi} \bar{\phi}' }+\frac{q^2 \bar{h} \bar{\Psi}\bar{\Psi}'}{r^2 \bar{\phi}'^2}\right]\delta w-\frac{\left(\frac{\bar{\Psi}}{r}\right)'}{1+\bar{f}_{R}}\delta u'-\left[ \frac{\left(\frac{\bar{\Psi}}{r}\right)'}{1+\bar{f}_{R}}\left(\frac{1}{r}+\frac{\bar{\xi}'}{\bar{\xi}}\right) + \frac{\left(\frac{\bar{\Psi}}{r}\right)''}{1+\bar{f}_{R}}-\right.\nonumber\\
&&\left.\frac{m_{0}^2\bar{\Psi}}{r \bar{N}(1+\bar{f_{R}})}\right]\delta u+\frac{\delta \phi''}{\bar{\phi}'} -\frac{\bar{\phi}''}{\bar{\phi}'^2}\delta \phi'=0.
\end{eqnarray}
As a result we find three perturbation equations (\ref{50}-\ref{52}) where the first two are dynamical equations and the third is the constraint \cite{ra}.

To get boundary conditions, perturbation modes need to satisfy ingoing boundary conditions near the event horizon
\begin{eqnarray}\label{53}
&&\delta u(t,r)=\mbox{Re}[ e^{-i\omega (t+r_{*})} \tilde{u}(r)],\nonumber \\
&&\delta w(t,r)=\mbox{Re}[ e^{-i\omega (t+r_{*})} \tilde{ w}(r)],\nonumber \\
&&\delta \phi(t,r)=\mbox{Re}[ e^{-i\omega (t+r_{*})} \tilde{\phi}(r)].
\end{eqnarray}
Near horizon, complex functions $\tilde{u}(r)$, $\tilde{ w}(r)$ and $\tilde{\phi}(r)$ have regular Taylor expansions
\begin{eqnarray}\label{54}
&&\tilde{u}(r)=\tilde{u}_{0}+\tilde{u}_{1}(r-r_{h})+\tilde{u}_{2} (r-r_{h})^2/2+...\nonumber \\
&&\tilde{w}(r)=\tilde{w}_{0}+\tilde{w}_{1}(r-r_{h})+\tilde{w}_{2} (r-r_{h})^2/2+...,\nonumber \\
&&\tilde {\phi}(r)=\tilde {\phi}_{0}+\tilde {\phi}_{1}(r-r_{h})+\tilde {\phi}_{2} (r-r_{h})^2/2+....
\end{eqnarray}
We get $\tilde{u}_{1}$, $\tilde{w}_{1}$ and $\tilde {\phi}_{1}$ in terms of $\tilde{u}_{0}$, $\tilde{ w}_{0}$ and $\omega$ by plugging (\ref{53}) and (\ref{54}) in (\ref{50}-\ref{52})
 \begin{eqnarray}\label{305a}
&&\tilde{\phi}_{1}=\frac{-q\psi_{h}{\omega}^{2}\left(\frac{{\phi'_{h}}^{2}}{1+f_{R_{0}}}+\frac{N'_{h} h_{h}}{r_{h}}\right)\tilde{w}_{0}+\left(\frac{\phi'_{h}N'_{h}\psi'_{h}h_{h}}{1+f_{R_{0}}}\left(\frac{i\omega}{\sqrt{h_{h}}}-N'_{h}\right)+\frac{\phi'_{h}N'_{h}m_{0}^{2}\psi_{h}h_{h}}{1+f_{R_{0}}}\right)\tilde{u}_{0}}{\omega\left(\omega+iN'_{h}\sqrt{h_{h}}\right)},\nonumber\\
&&\tilde{w}_{1}=\frac{\left[\frac{N'_{h}\sqrt{h_{h}}}{r_{h}}+m_{0}^{2}\sqrt{h_{h}}-\frac{iq^2\psi_{h}^{2}\omega}{N'_{h} \sqrt{h_{h}}}\left(\frac{r_{h}{\phi'_{h}}^{2}}{N'_{h}\sqrt{h_{h}}\left(1+f_{R_{0}}\right)}+\sqrt{h_{h}}\right)\right]\tilde{w}_{0}-\frac{q\phi'_{h}}{N'_{h}\sqrt{h_{h}}}\left(2+\frac{r_{h}\psi_{h}\psi'_{h}}{ 1+f_{R_{0}}}\right)\tilde{u}_{0}-\frac{i\omega q r_{h} \psi_{h}}{{N'_{h}}^{2}h_{h}}\tilde{\phi}_{1}}{N'_{h} \sqrt{h_{h}}-2i\omega},\nonumber \\
&&\tilde{u}_{1}=\frac{\left[\frac{N'_{h}\sqrt{h_{h}}}{r_{h}}-\frac{\sqrt{h_{h}}{\psi'_{h}}^{2}r_{h}N'_{h}}{1+f_{R_{0}}}+m_{0}^{2}\sqrt{h_{h}}\left(1+\frac{2r_{h}\psi_{h}\psi'_{h}}{1+f_{R_{0}}}\right)\right]\tilde{u}_{0}+\left(\frac{i\omega q \phi'_{h}}{N'_{h}}\left(-2N'_{h}+\frac{m_{0}^{2}r_{h}\psi_{h}^{2}}{1+f_{R_{0}}}\right)-\frac{2q\phi'_{h}{\omega}^{2}}{N'_{h}\sqrt{h_{h}}}\right)\tilde{w}_{0}}{N'_{h}\sqrt{h_{h}}-2i\omega} ,
\end{eqnarray}
where one fixes $\tilde{ w}_{0}=1$.

\end{document}